\documentclass[12pt]{article}
\usepackage{epsf}
\usepackage{amsmath,amssymb}

\usepackage[dvips]{graphicx}

\usepackage{comment}

\begin{document}

\newcommand{\lsim}{\stackrel{<}{_\sim}}
\newcommand{\gsim}{\stackrel{>}{_\sim}}

\newcommand{\rem}[1]{{$\spadesuit$\bf #1$\spadesuit$}}

\renewcommand{\theequation}{\thesection.\arabic{equation}}

\renewcommand{\thefootnote}{\fnsymbol{footnote}}
\setcounter{footnote}{0}

\begin{titlepage}

\def\thefootnote{\fnsymbol{footnote}}

\begin{center}

\hfill UT-16-15\\
\hfill April, 2016\\

\vskip .75in

{\Large \bf 

  Bottom-Tau Unification in Supersymmetric Model\\
  with Anomaly-Mediation\\

}

\vskip .75in

{\large 
So Chigusa and Takeo Moroi
}

\vskip 0.25in

\vskip 0.25in

{\em Department of Physics, University of Tokyo,
Tokyo 113-0033, Japan}

\end{center}
\vskip .5in

\begin{abstract}

  We study the Yukawa unification, in particular, the unification of
  the Yukawa coupling constants of $b$ and $\tau$, in the framework of
  supersymmetric (SUSY) model.  We concentrate on the model in which
  the SUSY breaking scalar masses are of the order of the gravitino
  mass while the gaugino masses originate from the effect of anomaly
  mediation and hence are one-loop suppressed relative to the
  gravitino mass.  We perform an accurate calculation of the Yukawa
  coupling constants of $b$ and $\tau$ at the grand unified theory
  (GUT) scale, including relevant renormalization group effects and
  threshold corrections.  In particular, we study the renormalization
  group effects, taking into account the mass splittings among
  sfermions, gauginos, and the standard model particles.  We found
  that the Yukawa coupling constant of $b$ at the GUT scale is about
  $70\ \%$ of that of $\tau$ if there is no hierarchy between the
  sfermion masses and the gravitino mass.  Our results suggest sizable
  threshold corrections to the Yukawa coupling constants at the GUT
  scale or significant suppressions of the sfermion masses relative to
  the gravitino mass.

\end{abstract}

\end{titlepage}

\renewcommand{\thepage}{\arabic{page}}
\setcounter{page}{1}
\renewcommand{\thefootnote}{\#\arabic{footnote}}
\setcounter{footnote}{0}

\section{Introduction}
\label{sec:intro}
\setcounter{equation}{0}

Supersymmetry (SUSY) provides attractive solutions to problems which
cannot be solved within the framework of the standard model (SM).  In
particular, the unification of the SM $SU(3)_C\times SU(2)_L\times
U(1)_Y$ gauge interactions, which is the prediction of the
grand unified theory (GUT), may be realized if the mass scale of SUSY
particles is of $O(1-10)\ {\rm TeV}$, because three gauge coupling
constants meet at $\sim 10^{16}\ {\rm GeV}$ using the renormalization
group equations (RGEs) of the minimal SUSY SM (MSSM) above $Q\sim
O(1-10)\ {\rm TeV}$ (with $Q$ being the renormalization scale).  In
addition, the lightest SUSY particle (LSP), if it is neutral, is a
good candidate of the dark matter.

Although SUSY SM is theoretically well-motivated, there is no
experimental evidence of the existence of SUSY particles at around
TeV.  On the contrary, the LHC is pushing up the possible mass scale
of SUSY particles.  For example, in the simplified scenario, colored
SUSY particles below $\sim 1 - 1.5\ {\rm TeV}$ are excluded
\cite{ATLAS-CONF-2015-062, Khachatryan:2016kdk}.  In addition, the
observed Higgs mass of $\sim 125\ {\rm GeV}$ \cite{Agashe:2014kda}
suggests that the mass scale of the stops is near $10\ {\rm TeV}$ or
higher (see, for example, \cite{Bagnaschi:2014rsa}) in order to
enhance the radiative correction to the SM-like Higgs mass
\cite{Okada:1990vk, Okada:1990gg, Ellis:1990nz, Ellis:1991zd,
  Haber:1990aw}.\footnote
{The stop masses of a few TeV can also explain the SM-like Higgs mass
  of $\sim 125\ {\rm GeV}$ if the stop-stop-Higgs tri-linear coupling
  constant is sizable \cite{Okada:1990gg}.}
Thus, it is important to consider the possibility that some of the
SUSY particles (in particular, stops) are much heavier than TeV scale.

One of the theoretically well-motivated scenarios with heavy sfermions
is so-called anomaly mediation SUSY breaking (AMSB)
\cite{Giudice:1998xp, Randall:1998uk} or pure gravity mediation (PGM)
\cite{Ibe:2006de, Ibe:2011aa, ArkaniHamed:2012gw}.  In such a
scenario, sfermion masses are generated by the effect of supergravity,
including direct K\"ahler interaction between the SM chiral multiplets
and SUSY breaking fields, while the gaugino masses arise from the
effect of anomaly mediation.  Then, the sfermion masses can become
$O(10)\ {\rm TeV}$ while the gaugino masses are one-loop suppressed
relative to the sfermion masses.  If all the SUSY breaking fields in
the hidden sector have gauge quantum numbers for hidden gauge group
responsible for the SUSY breaking, for example, such a framework
naturally shows up.

If we assume the AMSB/PGM scenario, we should consider if the
motivations of SUSY, in particular, the unification of the gauge
groups and the LSP dark matter, are still viable.  For the latter, it
has been pointed out that the LSP can be dark matter if the LSP is
neutral Wino \cite{Giudice:1998xp, Moroi:1999zb, Hisano:2006nn}, which
can be realized in a wide parameter region of the AMSB/PGM scenario.
In addition, if the mass scale of the sfermions are of $O(10)\ {\rm
  TeV}$, the gauge coupling constants of $SU(3)_C$, $SU(2)_L$, and
$U(1)_Y$ still meet at $\sim 10^{16}\ {\rm GeV}$, which may suggest
the successful GUT in the AMSB/PGM scenario.

As well as the gauge coupling unification, one important prediction of
GUT is the Yukawa coupling unification.  In particular, in a large
class of models (including simple GUTs based on $SU(5)$ gauge group),
$b$ and $\tau$ are embedded into a single multiplet of the GUT gauge
group, resulting in the unification of the Yukawa coupling constants
of $b$ and $\tau$.  Thus, it is important to check if the $b$-$\tau$
unification is viable in the AMSB/PGM scenario \cite{Tobe:2003bc}.  In
particular, it is necessary to consider the implication of the Higgs
mass constraint to the $b$-$\tau$ unification.  Before the discovery
of the Higgs boson, the $b$-$\tau$ unification has been already
studied for the case where the masses of SUSY particles are fairly
degenerate and are at the electroweak to TeV scale \cite{Ross:2007az,
  Antusch:2008tf, Antusch:2009gu}.  Then, in such a case, it has been
shown that, for a successful $b$-$\tau$ unification, relatively large
threshold corrections to the Yukawa coupling constant at the mass
scale of SUSY particles are suggested (assuming that the threshold
correction at the GUT scale is negligible).\footnote
{After the discovery of the Higgs boson, it has been discussed if the
  $b$-$\tau$ unification is successful in the so-called desert
  scenario in which the MSSM consistent with the observed Higgs mass
  of $\sim 125\ {\rm GeV}$ is valid up to the GUT scale.  For early
  attempts, see \cite{Baer:2012by, Baer:2012cp, Badziak:2012mm,
    Joshipura:2012sr, Elor:2012ig, Baer:2012jp, Anandakrishnan:2012tj,
    Ajaib:2014ana, Miller:2014jza, Gogoladze:2015tfa}.  Many of the
  studies consider, however, the cases where the mass scale of the
  SUSY particles are relatively close to the weak scale.}
We quantitatively study the $b$-$\tau$ unification with hierarchical
mass spectrum of the SUSY particles, accurately calculating the Yukawa
coupling constants of $b$ and $\tau$ at the GUT scale, taking into
account the mass splitting among SUSY particles.

In this paper, we study the Yukawa unification (in particular, the
$b$-$\tau$ unification) in SUSY $SU(5)$ GUT in the framework of the
AMSB/PGM scenario.  In such a scenario, as we have mentioned, there
are several important mass scales, i.e., the mass scale of sfermions,
that of gauginos, and the weak scale; at these scales, the particle
content of the relevant effective theory changes.  For an accurate
study of the $b$-$\tau$ unification, renormalization group effects
should be investigated taking proper effective theory at each scale.
In the past, $b$-$\tau$ unification was also studied for the cases
where the masses of SUSY particles are of $O(10)\ {\rm TeV}$, but the
effects of mass splitting among SUSY particles were taken into account
at the leading logarithmic level \cite{Baer:2012cp, Joshipura:2012sr,
  Elor:2012ig, Anandakrishnan:2012tj}.  Here, we solve the relevant
RGEs for each scale, include the threshold corrections, and study
$b$-$\tau$ unification in the framework of the AMSB/PGM scenario.

The organization of this paper is as follows.  In Section
\ref{sec:model}, we introduce the model we consider.  In Section
\ref{sec:numerical}, our numerical results are shown.  In particular,
we calculate the GUT scale values of the Yukawa coupling constants of
$b$ and $\tau$, and discuss how well they agree.  Implications of our
numerical results are discussed in Section \ref{sec:implications}.
The results are summarized in Section \ref{sec:summary}.

\section{Model: Brief Overview}
\label{sec:model}
\setcounter{equation}{0}

First, let us introduce the model we consider.  We consider the
AMSB/PGM scenario in which scalars as well as Higgsinos acquire masses
from direct couplings to the SUSY breaking fields while the gaugino
masses originate from the AMSB effect.  Then, we consider three
effective theories from the weak scale to the GUT scale.  We call
these effective theories as SM, $\tilde{G}$SM, and MSSM.  We consider
the case where the masses of the heavy Higgses are of the order of the
sfermion masses, and hence each effective theory consists of the
following particles:\footnote
{We assume that there is no new particle between $M_{\rm S}$ and
  $M_{\rm GUT}$ which significantly affects the renormalization group
  runnings of the MSSM parameters.}
\begin{itemize}
\item SM for $m_t<Q<M_{\tilde{G}}$: SM particles,
\item $\tilde{G}$SM for $M_{\tilde{G}}<Q<M_{\rm S}$: SM particles
    and gauginos,
\item MSSM for $M_{\rm S}<Q<M_{\rm GUT}$: MSSM particles,
\end{itemize}
where $M_{\rm S}$ is the mass scale of the sfermions, $M_{\tilde{G}}$
is the mass scale of gauginos, and $M_{\rm GUT}$ is the GUT scale
which is defined as the scale at which $U(1)_Y$ and $SU(2)_L$ gauge
coupling constants become equal.

In our study, the most important part of the superpotential is denoted
as\footnote
{For notational simplicity, we use same notations for the SM fields
  and the corresponding superfields.  In addition, the $SU(3)_C$ and
  $SU(2)_L$ indices are omitted.}
\begin{align}
  W = \mu H_u H_d + y_b H_d q_L b_R^{c} + y_\tau H_d l_L \tau_R^{c} 
  + y_t H_u q_L t_R^{c},
\end{align}
where $H_u$ and $H_d$ are up- and down-type Higgses, respectively,
while $q_L$, $t_R^{c}$, $b_R^{c}$, $l_L$, and $\tau_R^{c}$ are quarks
and leptons in third generation with $({\bf 3},{\bf 2},\frac{1}{6})$,
$({\bf \bar{3}},{\bf 1},-\frac{2}{3})$, $({\bf \bar{3}},{\bf
  1},\frac{1}{3})$, $({\bf 1},{\bf 2},-\frac{1}{2})$, and $({\bf
  1},{\bf 1},1)$ representations of $SU(3)_C\times SU(2)_L\times
U(1)_Y$ gauge group, respectively.  In addition, the relevant part of
the soft SUSY breaking terms are given by
\begin{align}
  {\cal L}_{\rm soft} = 
  - B_\mu H_u H_d
  - A_b H_d \tilde{q}_L \tilde{b}_R^{c}
  - A_t H_u \tilde{q}_L \tilde{t}_R^{c} 
  - \frac{1}{2} M_1 \tilde{B} \tilde{B}
  - \frac{1}{2} M_2 \tilde{W} \tilde{W}
  - \frac{1}{2} M_3 \tilde{g} \tilde{g}
  + \cdots ,
\end{align}
where $\tilde{B}$, $\tilde{W}$, and $\tilde{g}$ are Bino, Wino, and
gluino, respectively.  (The ``tilde'' is used for SUSY particles.)

Some of the Lagrangian parameters are related to each other at the GUT
scale.  For the soft SUSY breaking parameters, we neglect the
threshold corrections at the GUT scale.  Then, in $SU(5)$ GUT, we
parametrize the scalar masses at the GUT scale as
\begin{align}
  & m_{\tilde{Q}}^2 (M_{\rm GUT}) = m_{\tilde{U}}^2 (M_{\rm GUT}) =
  m_{\tilde{E}}^2 (M_{\rm GUT}) \equiv m_{\bf 10}^2,
  \\ &
  m_{\tilde{D}}^2 (M_{\rm GUT}) = m_{\tilde{L}}^2 (M_{\rm GUT}) \equiv
  m_{\bf \bar{5}}^2,
  \\ &
  m_{H_u}^2 (M_{\rm GUT}) \equiv m_{H{\bf 5}}^2,
  \\ &
  m_{H_d}^2 (M_{\rm GUT}) \equiv m_{H{\bf \bar{5}}}^2,
\end{align}
where $m_{\tilde{Q}}^2$, $m_{\tilde{U}}^2$, $m_{\tilde{D}}^2$,
$m_{\tilde{L}}^2$, and $m_{\tilde{E}}^2$ are soft SUSY breaking mass
squared parameters of the sfermions in $({\bf 3},{\bf
  2},\frac{1}{6})$, $({\bf \bar{3}},{\bf 1},-\frac{2}{3})$, $({\bf
  \bar{3}},{\bf 1},\frac{1}{3})$, $({\bf 1},{\bf 2},-\frac{1}{2})$,
and $({\bf 1},{\bf 1},1)$ representations of the SM gauge groups,
respectively, while $m_{H_u}^2$ and $m_{H_d}^2$ are those of $H_u$ and
$H_d$, respectively.  (For the sfermion masses, we assume the flavor
universality at the GUT scale for simplicity.)  In addition, the
gaugino masses arise from the AMSB effect, and are given by
\cite{Giudice:1998xp, Randall:1998uk}\footnote
{In the complete formula, the gaugino masses are proportional to the
  vacuum expectation value of the compensator field in supergravity.
  If the SUSY breaking field does not acquire vacuum expectation value
  as large as the Planck scale, however, the vacuum expectation value
  of the compensator field agrees with the gravitino mass.  In the
  following, we assume that is the case.}
\begin{align}
  M_1 (M_{\rm GUT}) =&\,  \frac{11g_1^{2} (M_{\rm GUT})}{16 \pi^{2}} m_{3/2},
  \\
  M_2 (M_{\rm GUT}) =&\,  \frac{g_2^{2} (M_{\rm GUT})}{16 \pi^{2}} m_{3/2},
  \\
  M_3 (M_{\rm GUT}) =&\,  -\frac{3g_3^{2} (M_{\rm GUT})}{16 \pi^{2}} m_{3/2},
\end{align}
where $g_1$, $g_2$, and $g_3$ are gauge coupling constants of
$U(1)_Y$, $SU(2)_L$, and $SU(3)_C$ gauge groups, respectively, and
$m_{3/2}$ is the gravitino mass which is taken to be a free parameter
in our analysis.  (We use the convention in which $m_{3/2}$ is real
and positive.)  The tri-linear scalar couplings also obey the AMSB
relation, and hence are one-loop suppressed relative to $m_{3/2}$.

At the mass scale of $Q=M_{\rm S}$, the Lagrangian parameters as well
as the fields in the MSSM are matched to those in the $\tilde{G}$SM.
The SM-like Higgs boson (which shows up at $Q<M_{\rm S}$) is given by
\begin{align}
  H_{\rm SM} = H_u \sin\beta + H_d^* \cos\beta,
\end{align}
with $\tan\beta$ being the ratio of the vacuum expectation values of
$H_u$ and $H_d$.  The Higgs potential of the $\tilde{G}$SM (and of the
SM) is expressed as
\begin{align}
  V_{\rm Higgs} = m_{H_{\rm SM}}^{2} H_{\rm SM}^{\dagger} H_{\rm SM} +
  \frac{\lambda}{2} (H_{\rm SM}^{\dagger} H_{\rm SM})^{2}.
\end{align}
The boundary condition for the quartic coupling constant is given by
\begin{eqnarray}
  \lambda (M_{\rm S}) = 
  \frac{g_1^{2} (M_{\rm S}) + g_2^{2} (M_{\rm S})}{4} \cos^{2} 2\beta
  + \delta \lambda,
\end{eqnarray}
where $\delta\lambda$ is the threshold correction due to the SUSY
particles (in particular, stops), which is taken into account in our
numerical calculation.  In addition, the mass of the pseudo-scalar
Higgs, which is embedded into the heavy Higgs multiplet, $H_{\rm
  heavy}=H_u \cos\beta - H_d^* \sin\beta$, is given by
\begin{align}
  m_A^2 = 
  \left[
    m_{H_u}^2 + m_{H_d}^2 + 2 \mu^2 - m_{H_{\rm SM}}^{2}
  \right]_{Q=M_{\rm S}}.
\end{align}

In our analysis, threshold corrections to the Yukawa coupling
constants play important role.  In particular, the correction to the
bottom Yukawa coupling may become sizable, and is studied by using the
parameter $\Delta_b$ with which the bottom Yukawa coupling constant
for $\tilde{G}$SM is given by
\begin{align}
  y_b^{(\tilde{G}{\rm SM})} (M_{\rm S}) =
  y_b (M_{\rm S}) \cos\beta (1 + \Delta_b).
\end{align}
The most important contributions to $\Delta_b$, which are proportional
to $\tan\beta$, come from the sbottom-gluino and stop-chargino
diagrams \cite{Hall:1993gn, Carena:1994bv, Blazek:1995nv}; at the
leading order of the mass-insertion approximation, $\Delta_b$ is given
by
\begin{align}
  \Delta_b \simeq
  \left[
    \frac{g_3^2}{6\pi^2} M_3 I(m_{\tilde b_1}^2,m_{\tilde b_2}^2,M_3^2)
    + \frac{y_t}{16\pi^2} A_t I(m_{\tilde t_1}^2,m_{\tilde t_2}^2,\mu^2)
  \right] \mu \tan\beta,
\end{align}
where $m_{\tilde b_1}$ and $m_{\tilde t_1}$ ($m_{\tilde b_2}$ and
$m_{\tilde t_2}$) are masses of lighter (heavier) stop and sbottom,
respectively, and
\begin{align}
  I(a,b,c) = -\frac{ab\ln (a/b)+bc\ln (b/c)+ca\ln (c/a)}{(a-b)(b-c)(c-a)}.
\end{align}
The unification of $y_b$ and $y_\tau$ crucially depends on $\Delta_b$.
Notice that, with large $\tan\beta$, the sign of $\Delta_b$ is
determined by ${\rm sign}(\mu)$.  We also note here that, when
$\tan\beta$ is not so large, other contributions to $\Delta_b$ may
become comparable to those from the sbottom-gluino and stop-chargino
loops.

For the calculation of the gaugino masses, we include the threshold
correction to the Wino and Bino masses from the Higgs-Higgsino loop
diagram \cite{Giudice:1998xp}:
\begin{align}
  \delta M_1 = \frac{g_1^{2} (M_{\rm S})}{16 \pi^{2}} L, ~~~
  \delta M_2 =&\,  \frac{g_2^{2} (M_{\rm S})}{16 \pi^{2}} L,
\end{align}
where
\begin{eqnarray}
 L \equiv \mu \sin2\beta 
  \frac{m_{A}^{2}}{\mu^{2}-m_{A}^{2}} \ln \frac{\mu^{2}}{m_{A}^{2}}.
  \label{L-parameter}
\end{eqnarray}

Then, at $Q=M_{\tilde{G}}$, the Lagrangian parameters in the
$\tilde{G}$SM are matched to those in the SM.  In particular, we
include the threshold correction to the gauge coupling constants from
the loop effects of gauginos.  The Lagrangian parameters at the weak
scale are related to those at $Q=M_{\tilde{G}}$ by using SM RGEs.
Then, the SM-like Higgs mass is evaluated as
\begin{eqnarray}
  m_h^2 = 2 \lambda (m_t) v^2 + \delta m_h^2,
\end{eqnarray}
where $v\simeq 174\ {\rm GeV}$ is the expectation value of the SM-like
Higgs boson and $\delta m_h^2$ is the threshold correction.

\section{Numerical Results}
\label{sec:numerical}
\setcounter{equation}{0}

Now, we perform the numerical calculation to study how well the
$b$-$\tau$ unification is realized in the AMSB/PGM scenario.  In
addition to the SM parameters, the present model contains seven new
parameters, $m_{\bf 10}^2$, $m_{\bf \bar{5}}^2$, $m_{H{\bf 5}}^2$,
$m_{H{\bf \bar{5}}}^2$, $m_{3/2}$, $\mu$, and $B_\mu$, with which the
Lagrangian parameters are determined.

Importantly, some of the Lagrangian parameters are determined by
low-energy observables, while boundary conditions for others are set
at the GUT scale as we have explained in the previous section.  In our
analysis, they are determined as follows:
\begin{itemize}
\item The gauge and Yukawa coupling constants are determined by using
  the data given in \cite{Agashe:2014kda}.  In particular, we use the
  bottom quark mass of $m_b (\overline{\rm MS})=4.18\ {\rm GeV}$, the
  top quark mass of $m_t=173.21\ {\rm GeV}$, and $\alpha_{3}
  (M_Z)=0.1185$ (with $\alpha_{3}=g_3^2/4\pi$).\footnote
  {We varied $m_t$ and $\alpha_{3}$ within the $1$-$\sigma$
    uncertainties, and checked that our conclusions are qualitatively
    unchanged.  In particular, the change of $R_{b\tau}$ given in Eq.\
    \eqref{Ratio} is at the level of a few $\%$.}
  Gauge and Yukawa coupling constants in the $\tilde{G}$SM and the
  MSSM are determined by taking into account the renormalization group
  runnings as well as relevant threshold corrections.
\item The soft SUSY breaking scalar mass squared parameters are fixed
  at the GUT scale.  (See the previous section.)  Some of them, as
  well as $\mu$ and $B_\mu$ parameters, are determined to fix the
  vacuum expectation value of the SM-like Higgs boson $v$,
  $\tan\beta$, and the Higgs mass $m_h$.  (For our numerical analysis,
  we use $m_h=125.09\ {\rm GeV}$ \cite{Agashe:2014kda}.)
\end{itemize}

With numerically solving RGEs, we determine sets of Lagrangian
parameters which are consistent with the low-energy and GUT scale
boundary conditions.  Our numerical calculation is based on the
SOFTSUSY package \cite{Allanach:2001kg}, in which three-loop RGEs for
the effective theory below the electoweak scale and two-loop RGEs
above $M_S$ are implemented.  We have implemented the three-loop RGEs
for the SM and the $\tilde{G}$SM, because those models are not
included in the original SOFTSUSY package.  (The RGEs for the SM can
be found in \cite{Buttazzo:2013uya}.  We have calculated the RGEs for
the $\tilde{G}$SM by taking into account the effects of gauginos.)  In
addition, one-loop threshold corrections due to the diagrams with SUSY
particles in the loop are included at relevant scales; those with only
gauginos in the loop are taken into account at $Q=M_{\tilde{G}}$,
while others at $Q=M_{\rm S}$.  In our numerical calculation, $M_{\rm
  S}$ is taken to be the geometric mean of the stop masses, while
$M_{\tilde{G}}=|M_3|$.  Following \cite{Buttazzo:2013uya}, we also
included two-loop threshold corrections to $\lambda$, $m_{H_{\rm
    SM}}^{2}$, $g_2$ and $g_1$ at $Q=m_t$, and two-loop plus
three-loop pure QCD corrections to $y_t$ and $g_3$.

With the boundary conditions which we adopt, $y_b$ and $y_\tau$ are
not guaranteed to be equal at the GUT scale, because the Yukawa
coupling constants are determined by using the fermion masses.  The
difference between $y_b(M_{\rm GUT})$ and $y_\tau (M_{\rm GUT})$
should be compensated by threshold corrections at the GUT scale if $b$
and $\tau$ are embedded into the same multiplet of the unified gauge
group; this is the case in simple SUSY GUT models based on $SU(5)$ (or
other unified gauge groups containing $SU(5)$).  To quantify the
$b$-$\tau$ unification, we define
\begin{align}
  R_{b\tau} = 
  \frac{y_b(M_{\rm GUT})}{y_\tau (M_{\rm GUT})}.
  \label{Ratio}
\end{align}
If the threshold correction at the GUT scale is negligible,
$R_{b\tau}$ should be close to unity.  We calculate $R_{b\tau}$ as a
function of model parameters, and study how it behaves.

\begin{figure}[t]
  \begin{tabular}{cc}
    \begin{minipage}[t]{0.49\hsize}
      \centering
      \includegraphics[width=90mm]{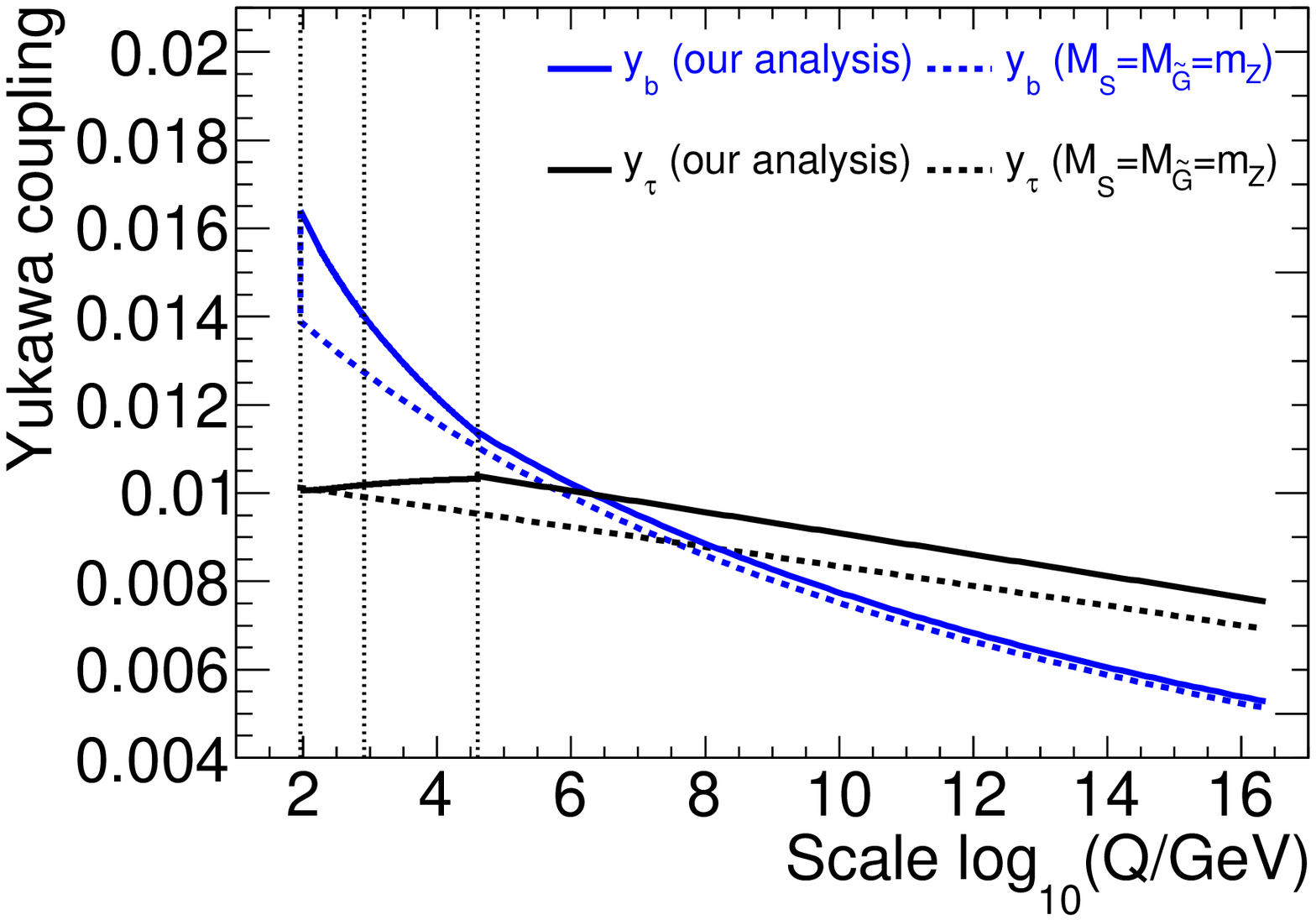}
    \end{minipage} &
    \begin{minipage}[t]{0.49\hsize}
      \centering
      \includegraphics[width=90mm]{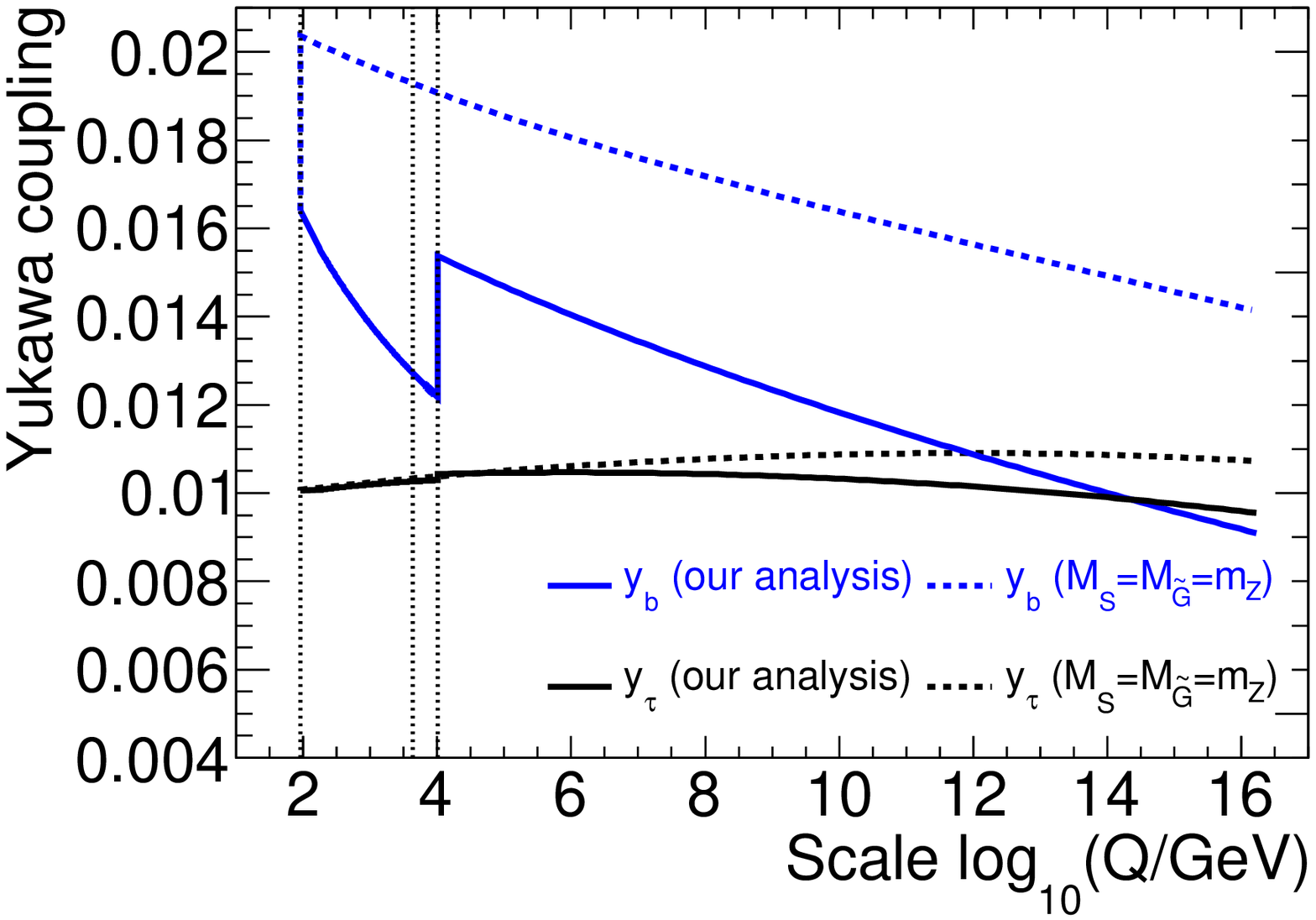}
    \end{minipage}\\
  \end{tabular}
  \caption{The renormalization group runnings of the Yukawa coupling
    constants of $b$ and $\tau$, taking $\tan\beta=3.95$, $m_{3/2}=50\
    {\rm TeV}$, $m_{\bf 10}=m_{\bf \bar{5}}=m_{3/2}$, $m_{H{\bf
        5}}=m_{H{\bf \bar{5}}}=0.8m_{3/2}$ and $\mu>0$ (left), and
    $\tan\beta=40.0,\ m_{3/2}=250\ {\rm TeV}$, $m_{\bf 10}=12\ {\rm
      TeV}$, $m_{\bf \bar{5}}=7\ {\rm TeV}$, $m_{H{\bf 5}}=m_{H{\bf
        \bar{5}}}=2\ {\rm TeV}$ and $\mu>0$ (right).  The solid lines
    are results with using the renormalization group analysis
    explained in Section \ref{sec:model}, while the dotted lines are
    the case where the theory is directly matched to the MSSM at
    $Q=m_Z$ (see the main text).  The vertical dotted lines are
    $Q=m_Z$, $M_{\tilde{G}}$, and $M_{S}$ to guide the eyes.  The
    ``jumps'' of the coupling constants at $Q=m_Z$, $M_{\tilde{G}}$,
    and $M_S$ are due to the threshold corrections.  The solid lines
    for $Q>M_S$ and the dotted lines for $Q>m_Z$ show the MSSM Yukawa
    coupling constants multiplied by $\cos\beta$, while the solid
    lines for $Q<M_S$ show the Yukawa coupling constants in the SM or
    $\tilde{G}$SM.}
  \label{fig:running}
\end{figure}

First, we show examples of the renormalization group runnings of the
Yukawa coupling constants from the electroweak scale to the GUT scale.
In Fig.\ \ref{fig:running}, we show how the Yukawa coupling constants
of $b$ and $\tau$ depend on the renormalization scale $Q$, taking
$\tan\beta=3.95$, $m_{3/2}=50\ {\rm TeV}$, $m_{\bf 10}=m_{\bf
  \bar{5}}=m_{3/2}$, $m_{H{\bf 5}}=m_{H{\bf \bar{5}}}=0.8m_{3/2}$ and
$\mu>0$ (left).  If the $b$-$\tau$ unification is studied by directly
matching the SM (after the electroweak symmetry breaking) to the MSSM
at $Q=m_Z$, some of the effects of the renormalization-group runnings
are not fully taken into account.  In addition, with such a procedure,
the effects of the wave function renormalization of the SM Higgs boson
on the running of the Yukawa coupling constants of $b$ and $\tau$ may
be nectlected.  The renormalization group runnings of the Yukawa
coupling constants with such an analysis are also shown in Fig.\
\ref{fig:running} to see the difference.  We can see that the
difference between the results of two analyses is sizable.  With the
present choice of parameters, we found that $R_{b\tau}\sim 0.7$ with
our analysis which properly takes into account the mass splittings
among MSSM particles, while $R_{b\tau}\sim 0.75$ with the analysis
taking $M_S=M_{\tilde{G}}=m_Z$.  In fact, we found that the difference
becomes larger if we take larger value of $\tan\beta$.  To see this,
we also show the renormalization group runnings, taking
$\tan\beta=40.0,\ m_{3/2}=250\ {\rm TeV}$, $m_{\bf 10}=12\ {\rm TeV}$,
$m_{\bf \bar{5}}=7\ {\rm TeV}$, $m_{H{\bf 5}}=m_{H{\bf \bar{5}}}=2\
{\rm TeV}$ and $\mu>0$ (right).  We can see a significant difference
between the two results.  This is due to the fact that, with large
$\tan\beta$, the threshold correction to $y_b$ at $Q=M_S$ becomes
large so that the GUT scale value of the Yukawa coupling constants
becomes sensitive what kind of RGEs are used between $m_Z\leq Q\leq
M_S$.  The $R_{b\tau}$ parameter gives important information about the
GUT scale values of the Yukawa coupling constants and the possible
size of the threshold corrections to the Yukawa coupling constants at
the GUT scale.  Thus, an accurate calculation of $R_{b\tau}$ is
important, for which, as we have seen, the use of proper effective
theory at each energy scale is needed.

To see how $R_{b\tau}$ depends on various model parameters, we
randomly choose $\sim 5\times 10^4$ sample points from the following
parameter space:
\begin{itemize}
\item $1.1 \leq \tan\beta \leq 60$,
\item $40 \, {\rm TeV} \leq m_{3/2} \leq 250 \, {\rm TeV}$,
\item $1 \, {\rm TeV} \leq m_X \leq 100 \, {\rm TeV}$ (with $X={\bf
    \bar{5}}$, ${H{\bf 5}}$, and ${H{\bf \bar{5}}}$),
\item $\mu >0$,
\end{itemize}
and calculate $R_{b\tau}$.\footnote
{We have accumulated more sample points for $m_{\bf 10},\ m_{\bf
    \bar{5}},\ m_{H{\bf 5}},\ m_{H{\bf \bar{5}}}<30\ {\rm TeV}$ than
  those with at least one scalar mass larger than $30\ {\rm TeV}$,
  because the sample points with small $|R_{b\tau}-1|$, which are of
  our interest, show up with relatively small scalar masses.  Thus,
  the density of the dots on the scatter plots has no meaning.}
In the AMSB/PGM scenario, the soft SUSY
breaking scalar mass squared parameters (i.e., $m_{\bf 10}^2$, $m_{\bf
  \bar{5}}^2$, $m_{H{\bf 5}}^2$, and $m_{H{\bf \bar{5}}}^2$ in the
present set up) are expected to be of $O(m_{3/2}^2)$.  However, we
also study the parameter regions where scalar masses and $m_{3/2}$ are
hierarchical.\footnote
{Our calculation becomes invalid when the scalar masses are much
  smaller than the gaugino masses.  For most of the sample points we
  studied, we have checked that the scalar masses are comparable to or
  larger than the gaugino masses.  This is partly because of the
  renormalization group effects due to gaugino masses.}
As we will see in the following, the sign of the $\mu$-parameter is
preferred to be positive to make $R_{b\tau}$ close to $1$.  Thus, the
scan is performed only in the parameter space with $\mu >0$.

\begin{figure}[t]
  \begin{tabular}{cc}
    \begin{minipage}[t]{0.49\hsize}
      \centering
      \includegraphics[width=90mm]{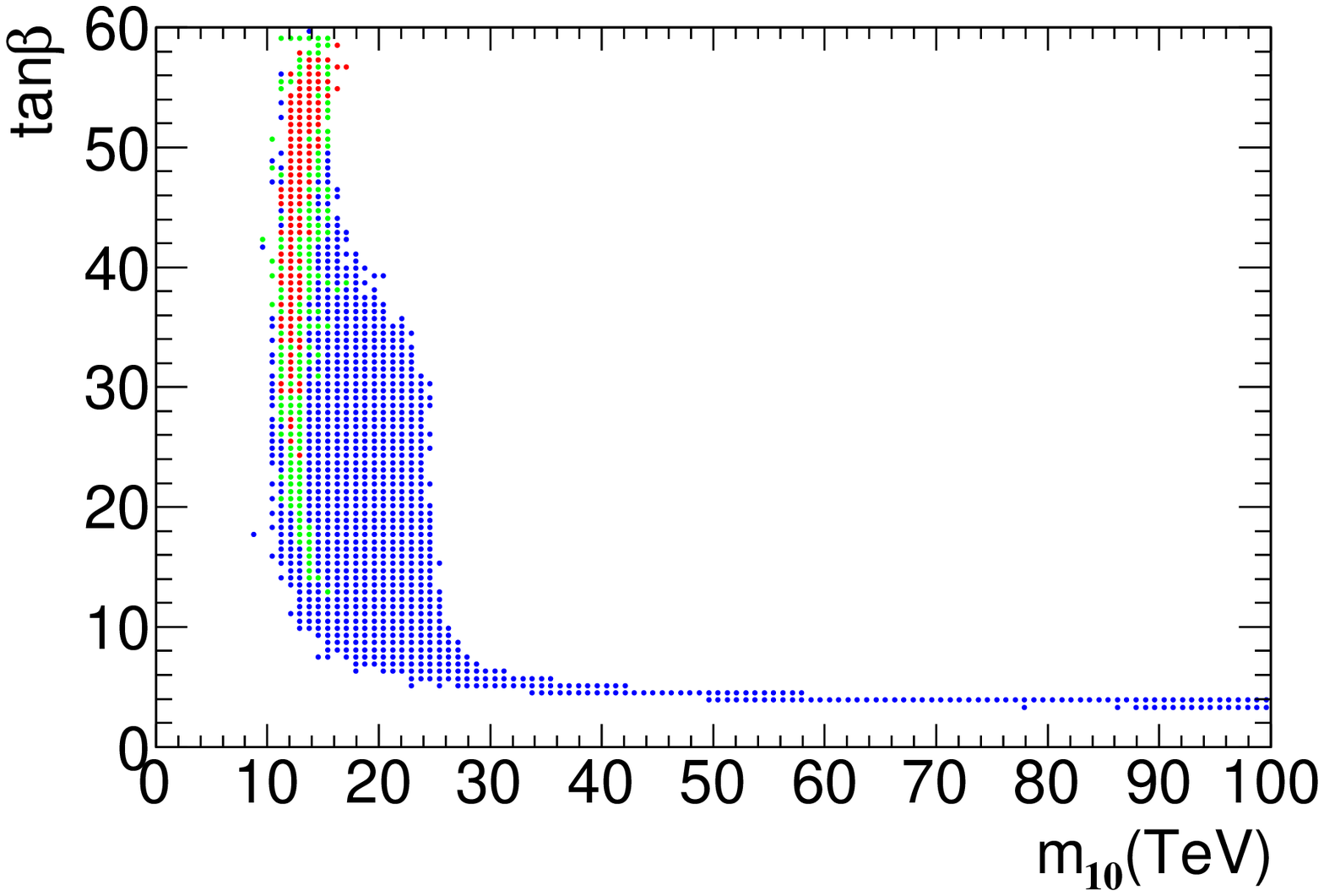}
    \end{minipage} &
    \begin{minipage}[t]{0.49\hsize}
      \centering
      \includegraphics[width=90mm]{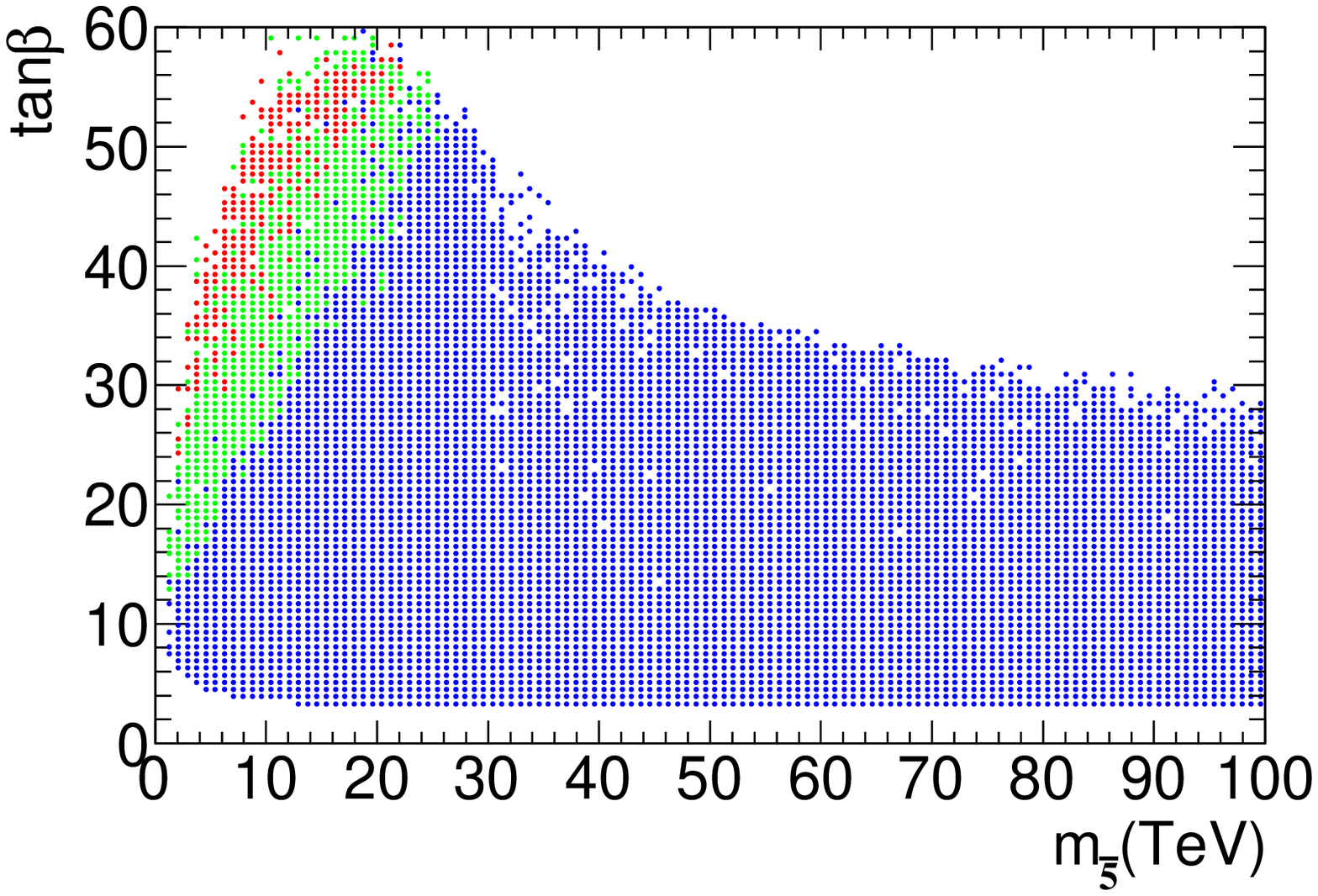}
    \end{minipage}\\
    \begin{minipage}[t]{0.49\hsize}
      \centering
      \includegraphics[width=90mm]{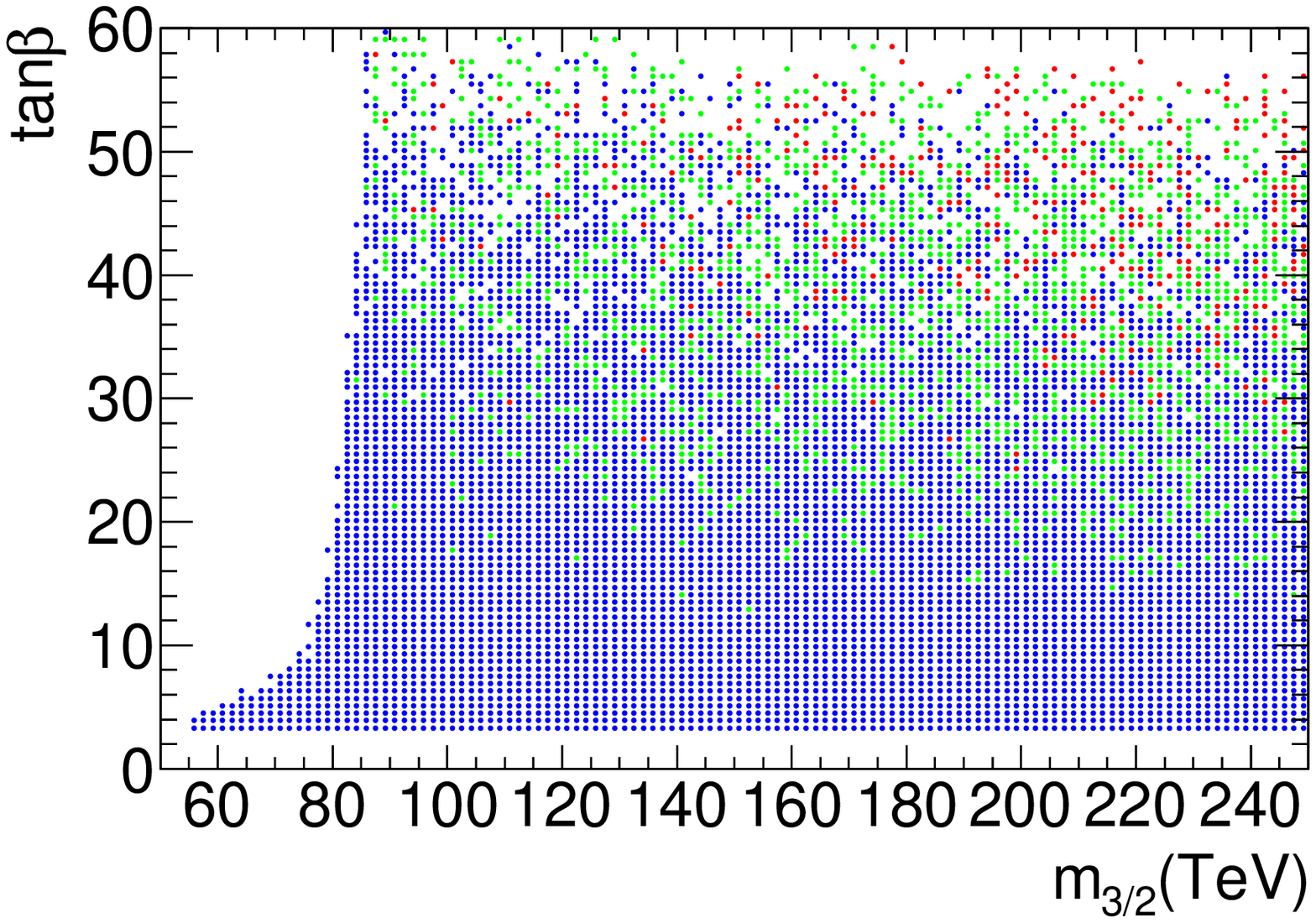}
    \end{minipage} &
    \begin{minipage}[t]{0.49\hsize}
      \centering
      \includegraphics[width=90mm]{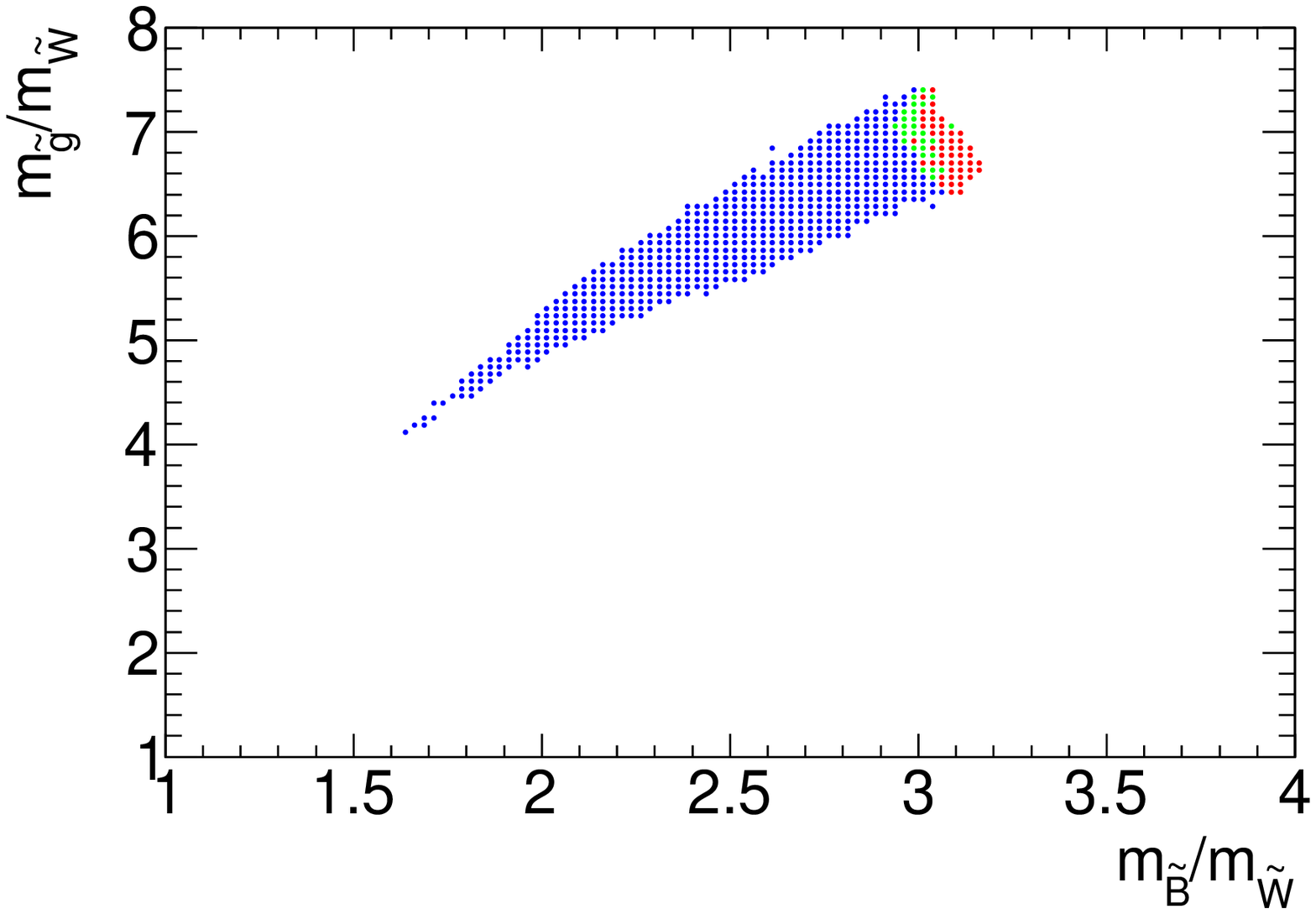}
    \end{minipage}\\
  \end{tabular}
  \caption{The distribution of the sample points on $m_{\bf 10}$ vs.\
    $\tan\beta$, $m_{\bf \bar{5}}$ vs.\ $\tan\beta$, $m_{3/2}$ vs.\
    $\tan\beta$, and $m_{\tilde{B}}/m_{\tilde{W}}$ vs.\
    $m_{\tilde{g}}/m_{\tilde{W}}$ planes.  The colors of the dots
    indicate the smallest value of $|R_{b\tau}-1|$: $|R_{b\tau}-1| <
    0.1$ (red), $0.1 < |R_{b\tau}-1| < 0.2$ (green), and
    $|R_{b\tau}-1| > 0.2$ (blue).}
  \label{fig:scan}
\end{figure}

In Fig.\ \ref{fig:scan}, we show the distribution of the sample points
we studied on $m_{\bf 10}$ vs.\ $\tan\beta$, $m_{\bf \bar{5}}$ vs.\
$\tan\beta$, $m_{3/2}$ vs.\ $\tan\beta$, and
$m_{\tilde{B}}/m_{\tilde{W}}$ vs.\ $m_{\tilde{g}}/m_{\tilde{W}}$
planes (with $m_{\tilde{B}}$, $m_{\tilde{W}}$, and $m_{\tilde{g}}$
being the on-shell masses of Bino, Wino, and gluino, respectively).
Here, we divide each plane into $120\times 100$ grids.  Then, if at
least one sample point falls into the grid, we put a dot on the grid.
The colors of the dots indicate the smallest value of $|R_{b\tau}-1|$
we found: $|R_{b\tau}-1| < 0.1$ (red), $0.1 < |R_{b\tau}-1| < 0.2$
(green), and $|R_{b\tau}-1| > 0.2$ (blue).  (Thus, the dots do not
represent the sample points.)

From the plot on the $m_{\tilde{B}}/m_{\tilde{W}}$ vs.\
$m_{\tilde{g}}/m_{\tilde{W}}$ plane, we can see that the Wino becomes
the lightest among the gauginos, and hence the Wino-like neutralino
becomes the LSP in the parameter space we studied.  We have imposed
the following experimental constraints on the gaugino masses from the
direct searches of gluino and long-lived Wino:\footnote
{If Wino-like neutralino is the LSP, and also if it is the dominant
  component of dark matter, there also exist cosmological and
  astrophysical constraints, like those from big-bang nucleothynthesis
  \cite{Kawasaki:2015yya}, $\gamma$-ray flux from Milky Way satellites
  \cite{Bhattacherjee:2014dya}, and anti-proton flux in the cosmic ray
  \cite{Giesen:2015ufa, Jin:2015sqa, Evoli:2015vaa, Ibe:2015tma,
    Hamaguchi:2015wga, Lin:2015taa}.  Such constraints can be,
  however, avoided if the Wino is not the dominant component of dark
  matter.}
\begin{itemize}
\item $m_{\tilde{g}} > 1.5\ {\rm TeV}$ \cite{ATLAS-CONF-2015-062,
    Khachatryan:2016kdk},
\item $m_{\tilde{W}} > 270\ {\rm GeV}$ \cite{Aad:2013yna, CMS:2014gxa}.
\end{itemize}
We show only the points consistent with the above constraints.  Notice
that, in the present model, all the sfermion masses are multi-TeV or
larger so that the experimental bounds on them are unimportant.

From the scatter plot on the $m_{\bf 10}$ vs.\ $\tan\beta$ plane, one
can see that $m_{\bf 10}$ and $\tan\beta$ are strongly correlated.
This is because the lightest Higgs mass $m_h$ is mostly determined by
these two parameters.  In the MSSM, the lightest Higgs mass is
predicted to be smaller than $m_Z|\cos 2\beta|$ at the tree level, and
a sizable radiative correction is necessary to push the Higgs mass up
to $\sim 125\ {\rm GeV}$.  In general, there are two important sources
of the radiative correction; one is the renormalization-group running
of the quartic Higgs coupling from $M_{\rm S}$ to the weak scale, and
the other is the threshold correction at $Q=M_{\rm S}$ due to
stop-stop-Higgs tri-linear coupling constant.  In the present model,
the tri-linear coupling is one-loop suppressed so that the latter
effect is insignificant.  Consequently, the Higgs mass is mostly
determined by the stop masses (which are determined by $m_{\bf 10}$)
and $\tan\beta$; in particular, larger values of the stop masses are
required to realize $m_h\simeq 125\ {\rm GeV}$ as $\tan\beta$
decreases.  As a result, there are two regimes in the parameter space.
One is with $m_{\bf 10}\lesssim 25\ {\rm TeV}$, resulting in
hierarchical masses $(m_{\bf 10}/m_{3/2})^2\lesssim 10^{-2}$ and
large $\tan\beta\gtrsim 10$; in such a region, $R_{b\tau}$ can be
close to the unity.  The other is with $m_{\bf 10}\gtrsim 25\ {\rm
  TeV}$, where $m_{\bf 10}$ can be of the same order of $m_{3/2}$ and
$\tan\beta$ becomes $\sim O(1)$; in such a region, $R_{b\tau}$ is
suppressed to be $\sim 0.7$.

\begin{figure}
  \begin{tabular}{cc}
    \begin{minipage}[t]{0.49\hsize}
      \centering
      \includegraphics[width=90mm]{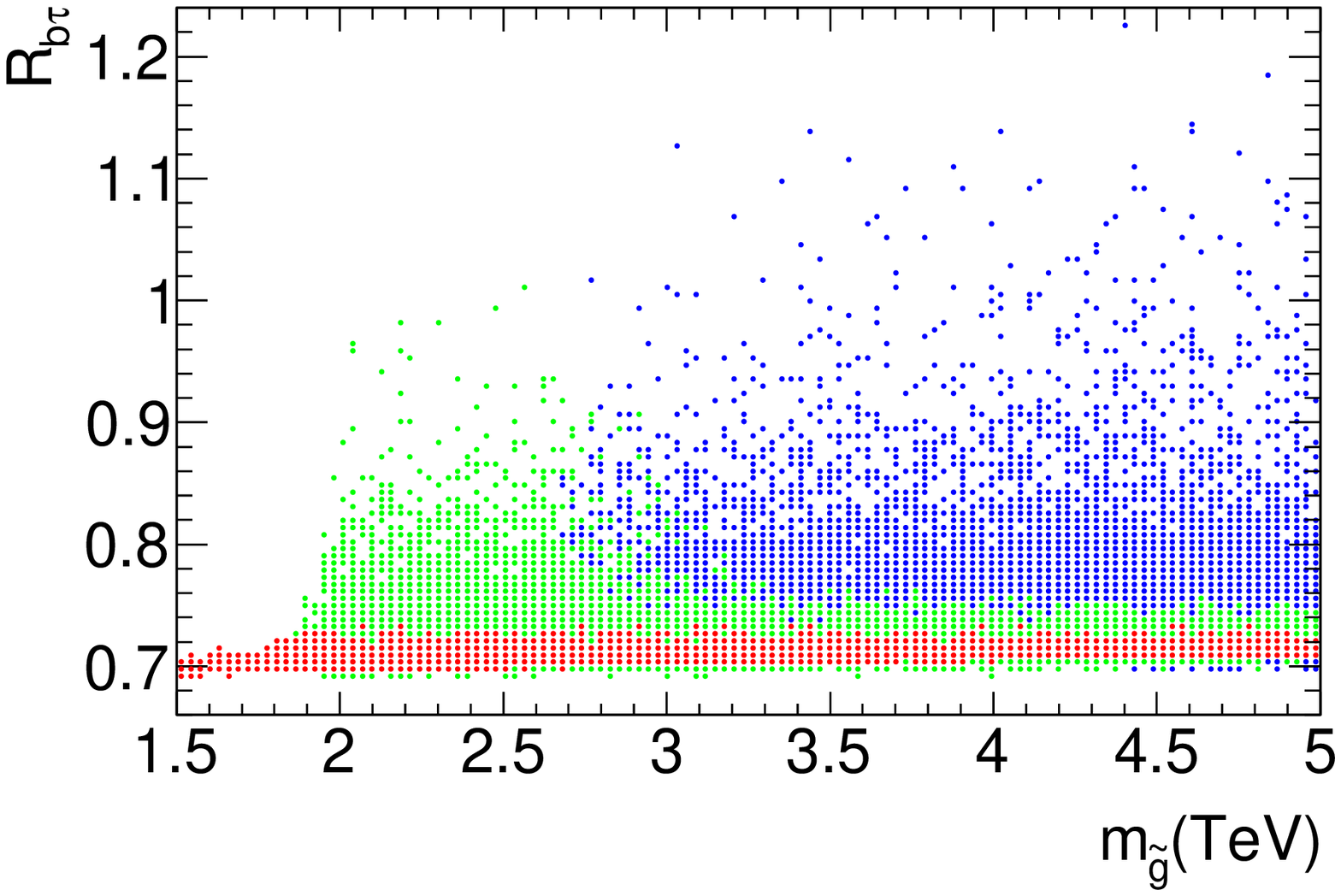}
    \end{minipage} &
    \begin{minipage}[t]{0.49\hsize}
      \centering
      \includegraphics[width=90mm]{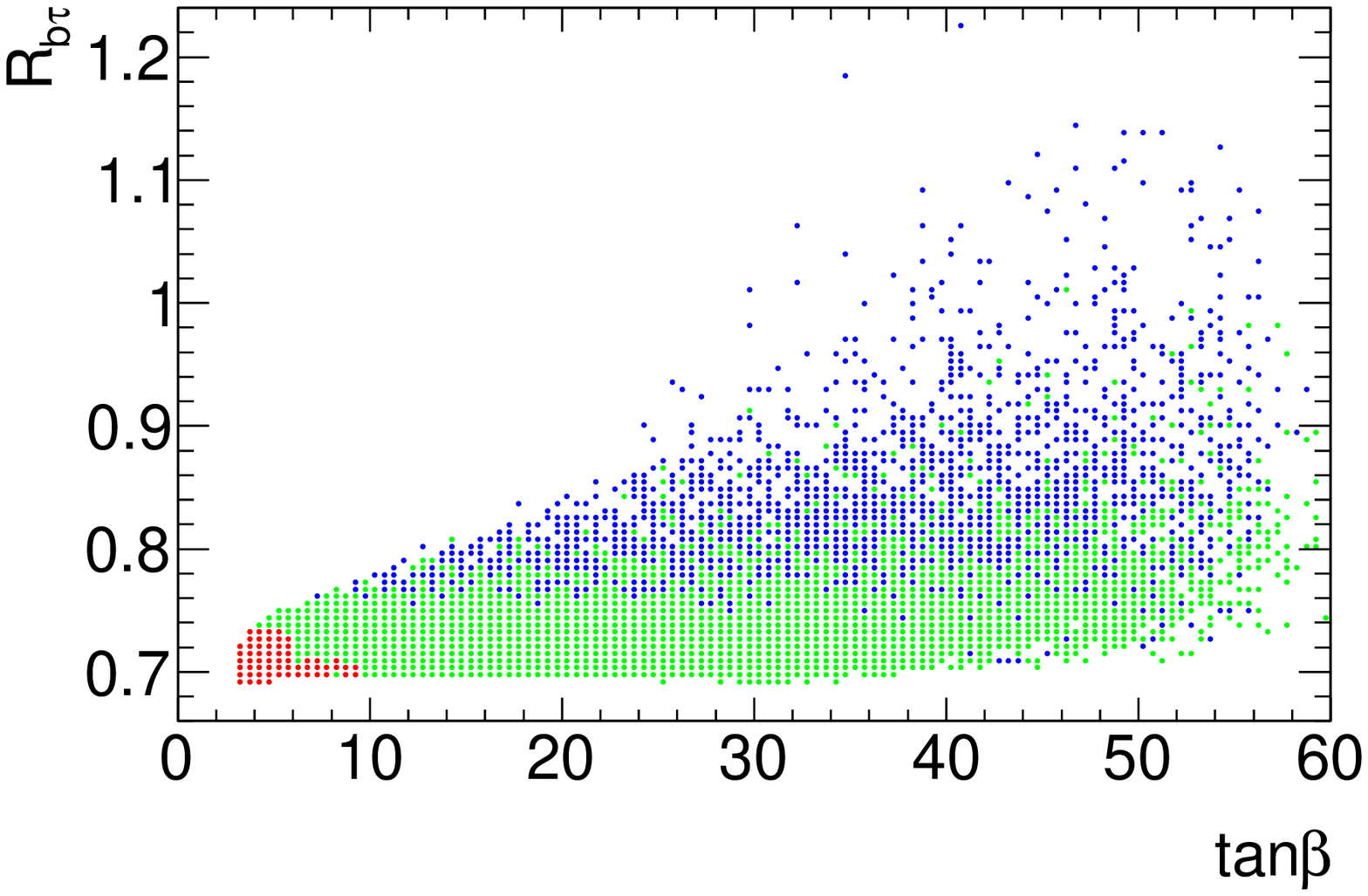}
    \end{minipage}\\
    \begin{minipage}[t]{0.49\hsize}
      \centering
      \includegraphics[width=90mm]{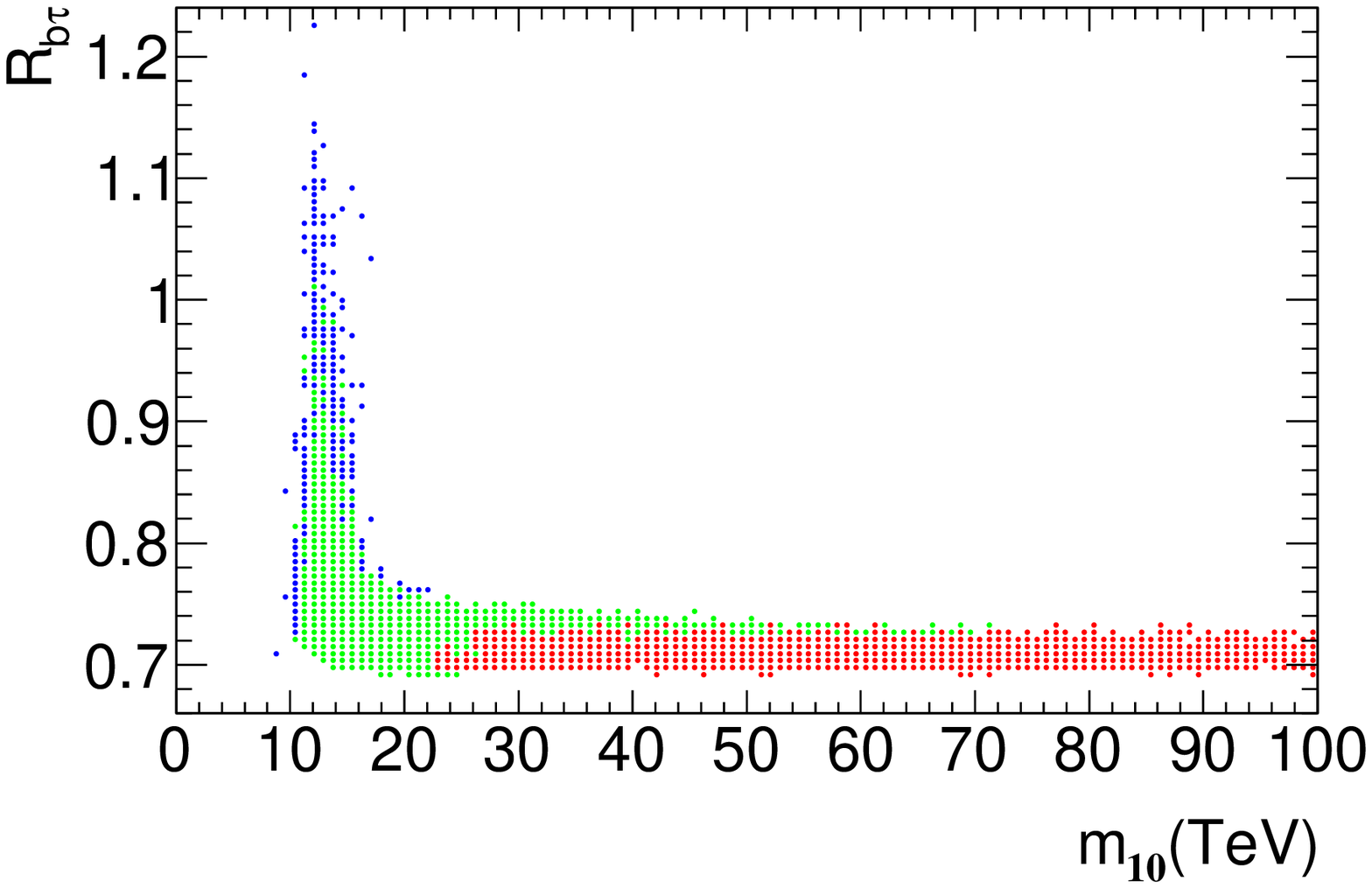}
    \end{minipage} &
    \begin{minipage}[t]{0.49\hsize}
      \centering
      \includegraphics[width=90mm]{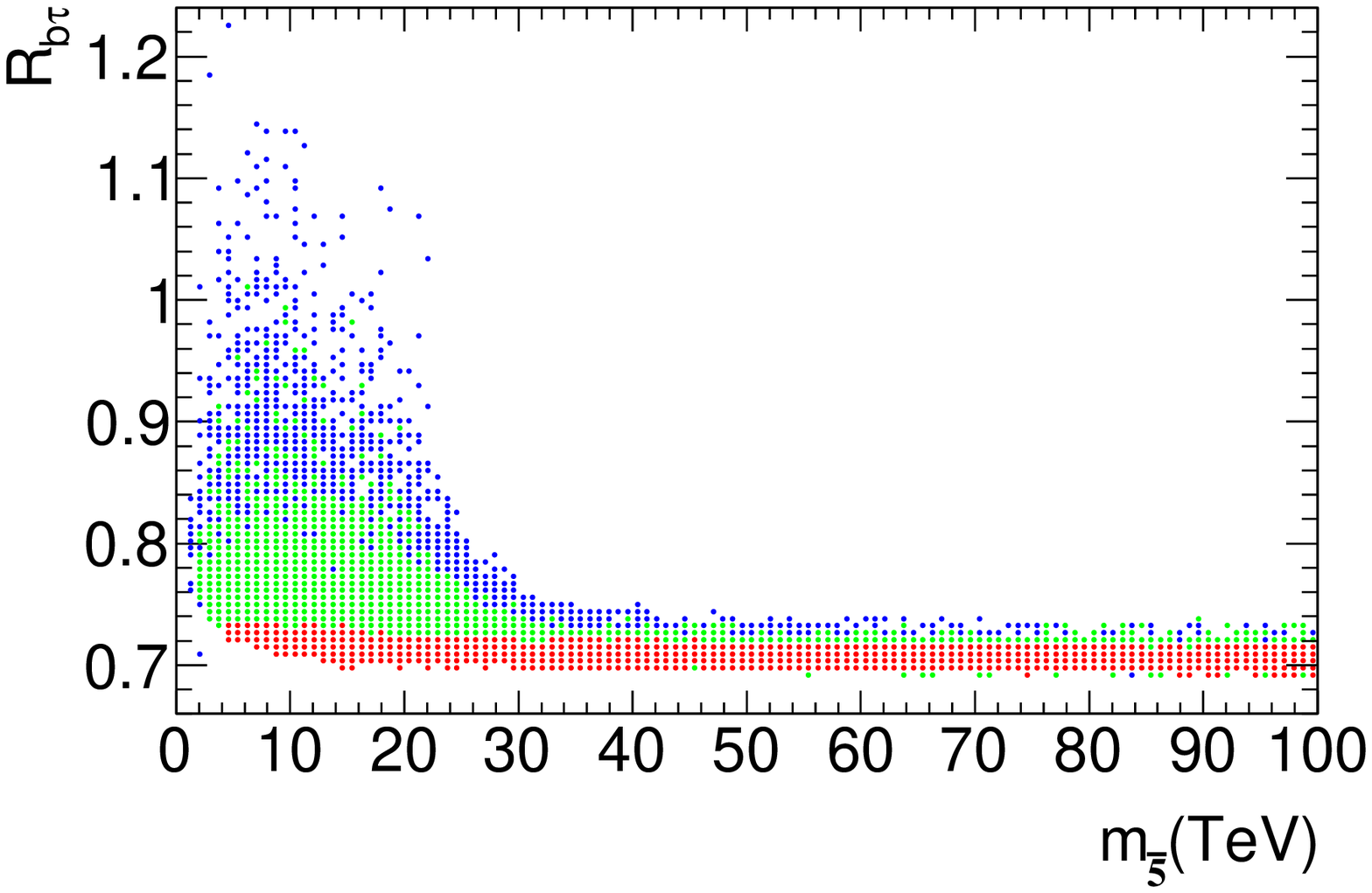}
    \end{minipage}\\
    \begin{minipage}[t]{0.49\hsize}
      \centering
      \includegraphics[width=90mm]{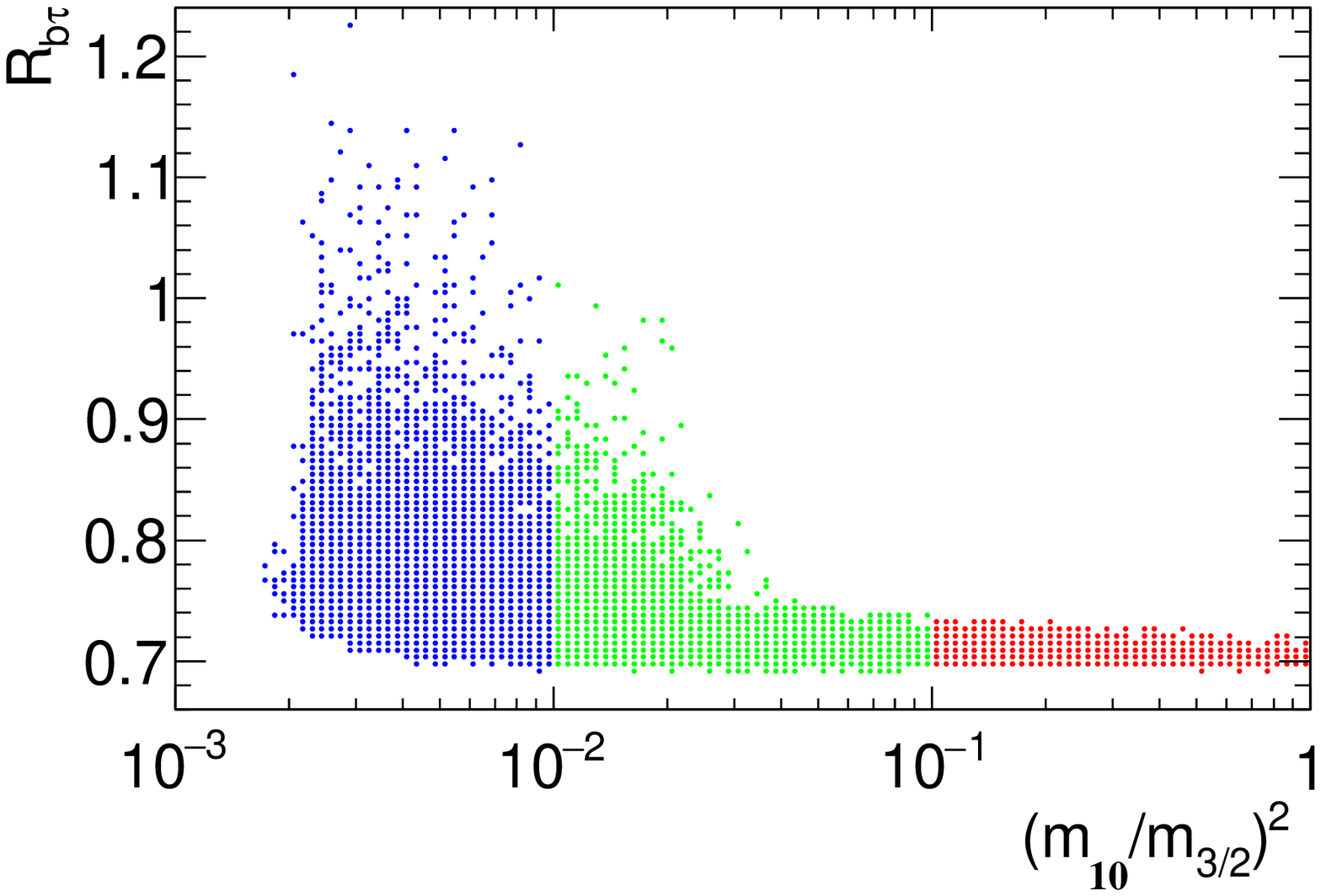}
    \end{minipage} &
    \begin{minipage}[t]{0.49\hsize}
      \centering
      \includegraphics[width=90mm]{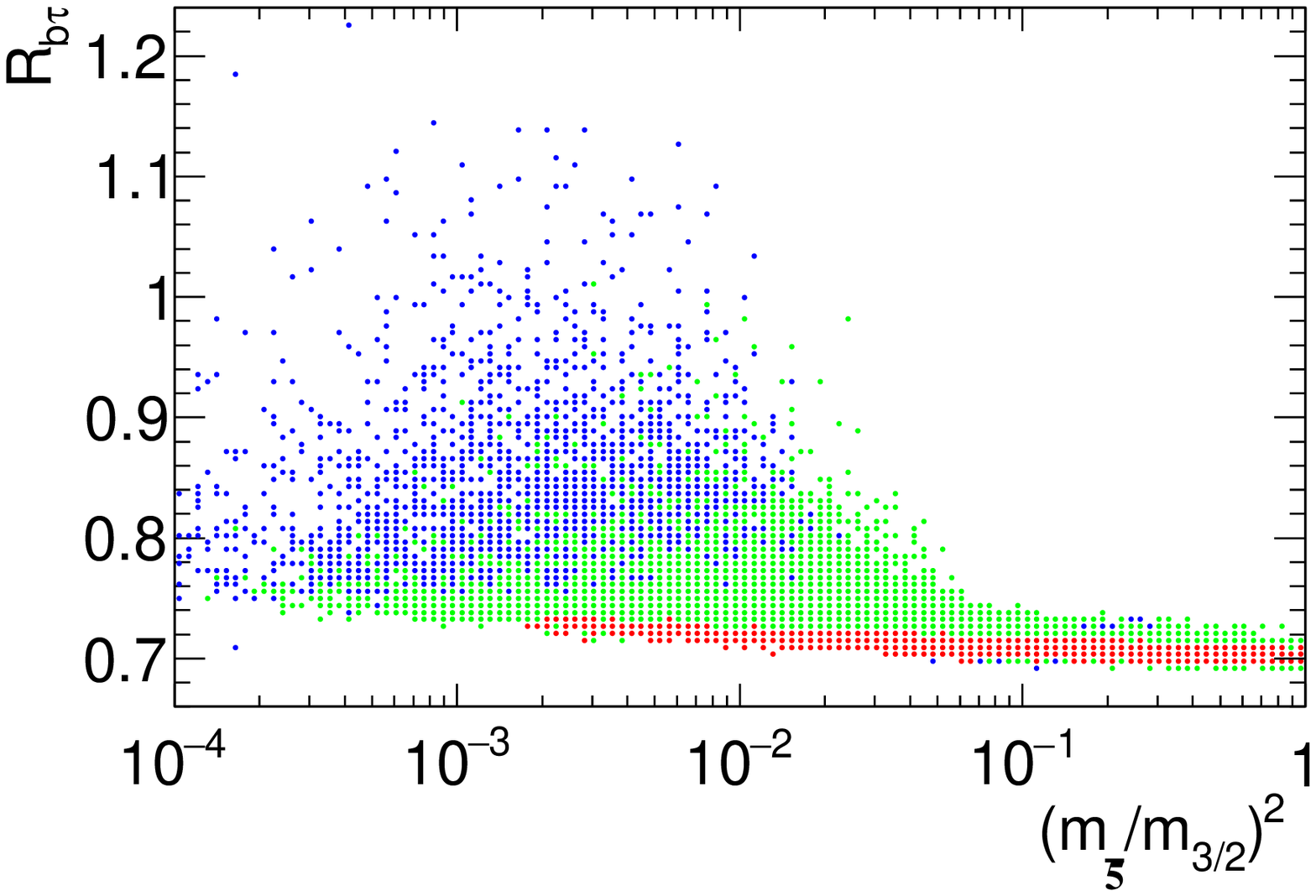}
    \end{minipage}\\
  \end{tabular}
  \caption{The distribution of the sample points on $m_{\tilde{g}}$
    vs.\ $R_{b\tau}$, $\tan\beta$ vs.\ $R_{b\tau}$, $m_{\bf 10}$ vs.\
    $R_{b\tau}$, $m_{\bf \bar{5}}$ vs.\ $R_{b\tau}$, $(m_{\bf 10} /
    m_{3/2})^2$ vs.\ $R_{b\tau}$, and $(m_{\bf \bar{5}} / m_{3/2})^2$
    vs.\ $R_{b\tau}$ planes.  The colors of the dots show the largest
    value of $(m_{\bf 10} / m_{3/2})^{2}$: $(m_{\bf 10} / m_{3/2})^{2}
    > 0.1$ (red), $0.01 < (m_{\bf 10} / m_{3/2})^{2} < 0.1$ (green),
    and $(m_{\bf 10} / m_{3/2})^{2} < 0.01$ (blue).}
  \label{fig:Rbtau}
\end{figure}

In Fig.\ \ref{fig:Rbtau}, we show the result of our random scan on
$m_{\tilde{g}}$ vs.\ $R_{b\tau}$, $\tan\beta$ vs.\ $R_{b\tau}$,
$m_{\bf 10}$ vs.\ $R_{b\tau}$, $m_{\bf \bar{5}}$ vs.\ $R_{b\tau}$,
$(m_{\bf 10} / m_{3/2})^2$ vs.\ $R_{b\tau}$, and $(m_{\bf \bar{5}} /
m_{3/2})^2$ vs.\ $R_{b\tau}$ planes.  As Fig.\ \ref{fig:scan}, we
divide the planes into grids, and put a dot on the grid if there is at
least one sample point falling into the grid.  The colors of the dots
show the largest value of $(m_{\bf 10} / m_{3/2})^{2}$ we found:
$(m_{\bf 10} / m_{3/2})^{2} > 0.1$ (red), $0.01 < (m_{\bf 10} /
m_{3/2})^{2} < 0.1$ (green), and $(m_{\bf 10} / m_{3/2})^{2} < 0.01$
(blue).  We notice here that, on the $\tan\beta$ vs.\ $R_{b\tau}$
plane, the dots exist only for $\tan\beta\gtrsim 3$.  This is because
the scan is restricted to the parameter region of $m_{\bf 10}< 100\
{\rm TeV}$, and hence the stop mass is at most $\sim 100\ {\rm TeV}$.
If a larger value of $m_{\bf 10}$ is considered, smaller value of
$\tan\beta$ is allowed.

From Fig.\ \ref{fig:Rbtau}, it is suggested that, in order to make
$R_{b\tau}$ close to $1$, (i) the scalar masses should be suppressed
compared to $m_{3/2}$, (ii) $\tan\beta$ should be large, and (iii)
$\mu>0$.  This is because $R_{b\tau}$ becomes $\sim 0.7$ when the
threshold correction $\Delta_b$ is negligible.  In particular, we
found that the renormalization-group effect between the weak scale and
$M_{\rm S}$ significantly suppresses $y_b$, which makes $R_{b\tau}$
smaller.  The conditions (i) and (ii) are necessary to enhance
$\Delta_b$.  In addition, the condition (iii) is necessary to make
$\Delta_b$ negative.  However, the condition (i) may conflict with the
simple expectation from the AMSB/PGM scenario which requires the soft
SUSY breaking scalar masses to be of $O(m_{3/2})$ \cite{Tobe:2003bc}.
The condition (ii), combined with $m_h\simeq 125\ {\rm GeV}$, suggests
that $m_{\bf 10}\sim 10\ {\rm TeV}$, as can be seen in the plot on the
$m_{\bf 10}$ vs.\ $\tan\beta$ plane of Fig.\ \ref{fig:scan}.  From the
plot on the $m_{\bf 5}$ vs.\ $\tan\beta$ plane, one can also see that
the sample points with small $|R_{b\tau}-1|$ concentrate on the region
with relatively small $m_{\bf \bar{5}}$.  This is because
$m_{H_u}^{2}(M_{\rm S})$ decreases for decreasing $m_{\bf \bar{5}}$
due to the renormalization group effect, and the small
$m_{H_u}^{2}(M_{\rm S})$ enhances $\mu^{2}$ determined by the
electroweak symmetry breaking condition. Since $\Delta_b$ is
proportional to $\mu$, the small $m_{\bf \bar{5}}$ is favored.

\begin{figure}[t]
  \begin{minipage}{0.48\hsize}
    \begin{center}
      \includegraphics[width=85mm]{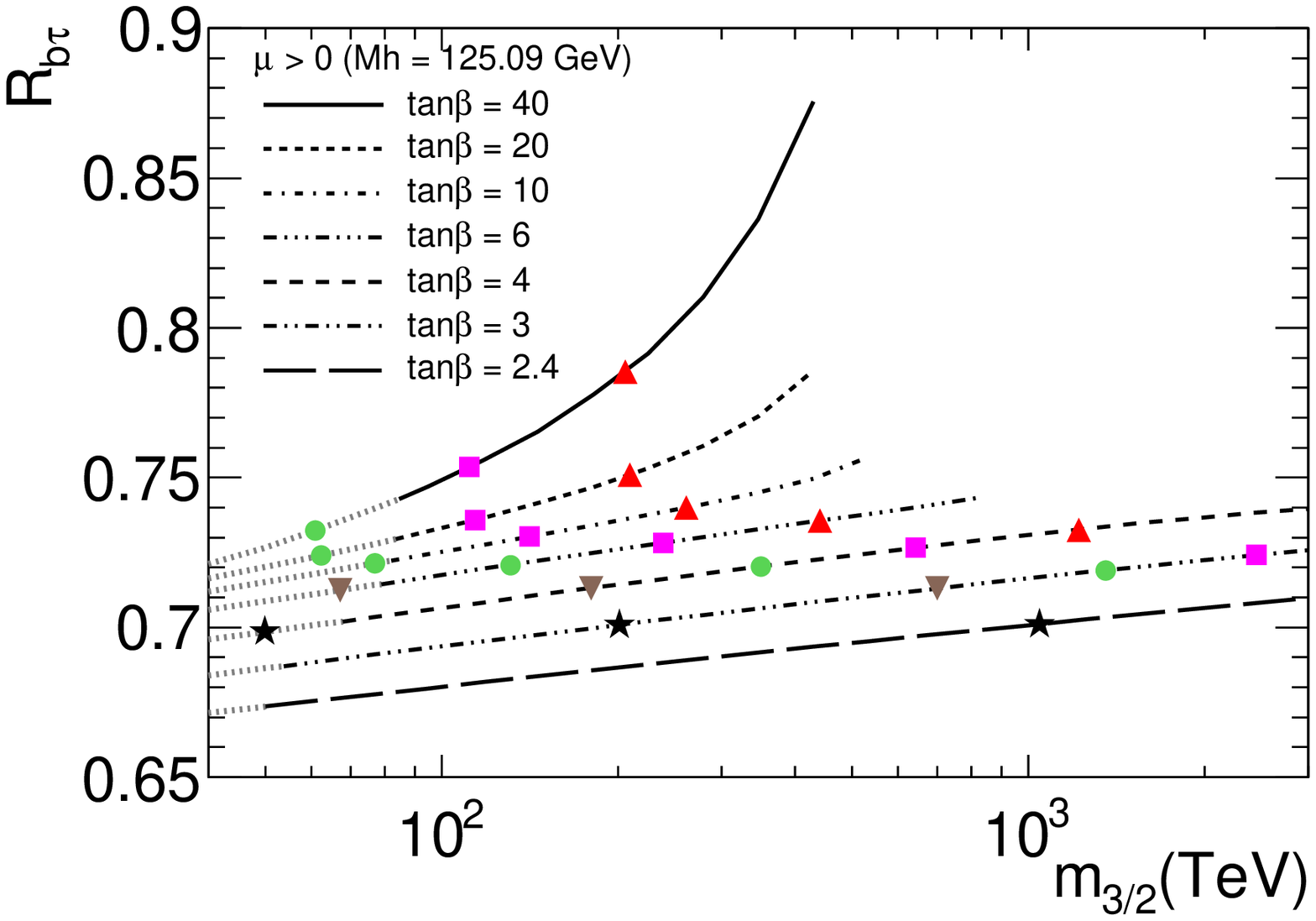}
    \end{center}
  \end{minipage}
  \begin{minipage}{0.48\hsize}
    \begin{center}
      \includegraphics[width=85mm]{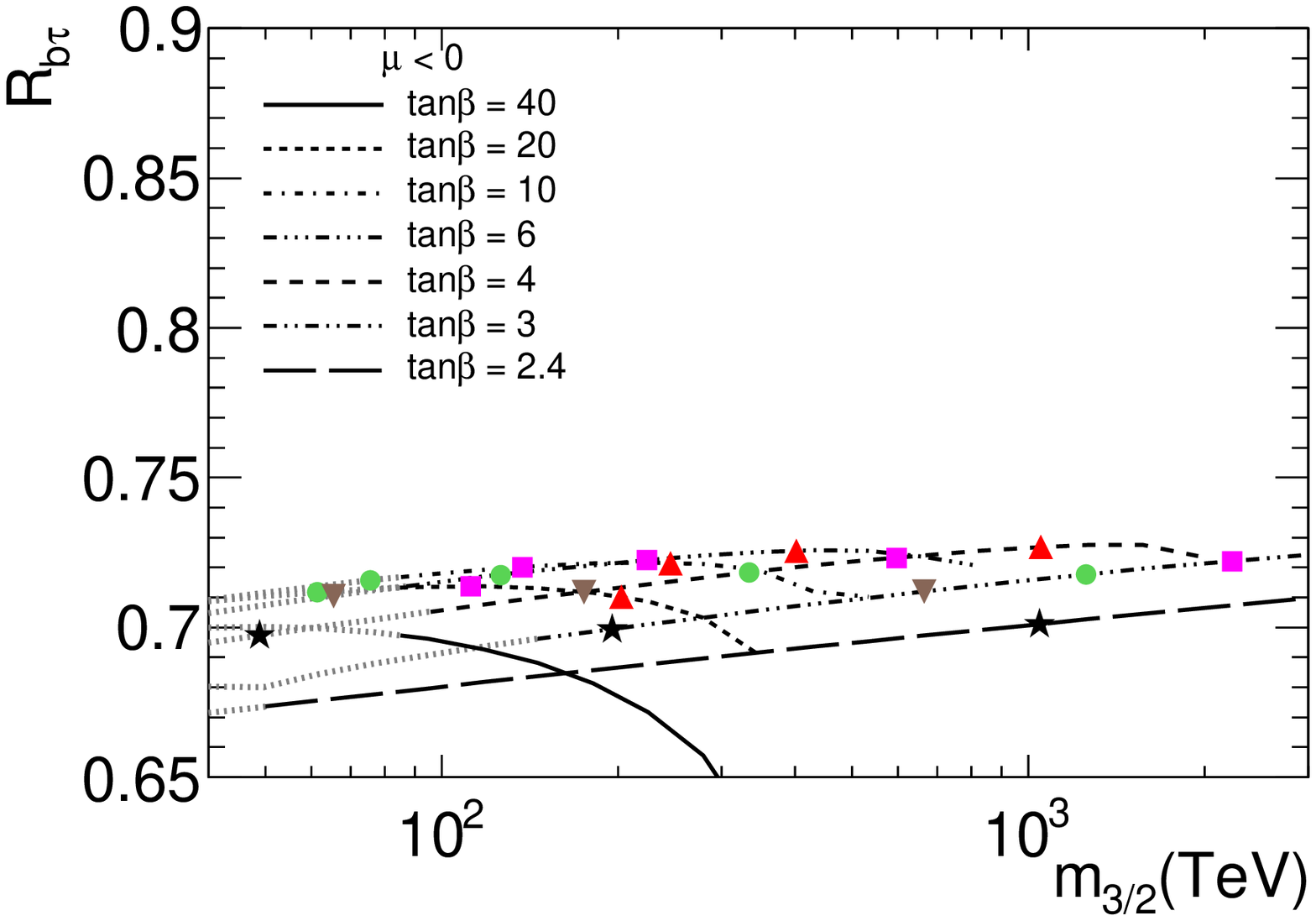}
    \end{center}
  \end{minipage}
  \caption{$R_{b\tau}$ as a function of $m_{3/2}$, taking $m_{\bf
      \bar{5}} = m_{\bf 10}$, and $m_{H {\bf 5}} = m_{H {\bf \bar{5}}}
    = 0.8m_{\bf 10}$ for $\mu >0$ (left), $\mu <0$ (right).  The value
    of $\tan\beta$ is taken to be $2.4$, $3$, $4$, $6$, $10$, $20$,
    and $40$ as shown in the figure.  The light gray (dotted) part of
    the lines correspond to the regions excluded by the Wino or gluino
    mass bounds.  The marks on the lines show the points with $(m_{\bf
      10} / m_{3/2})^{2}$ = 1.0($\bigstar$),
    0.1($\blacktriangledown$), 0.03({\Large $\bullet $}),
    0.01($\blacksquare$), and 0.003($\blacktriangle$).}
  \label{fig:MgravVsR}
\end{figure}

\begin{figure}[t]
  \begin{center}
    \includegraphics[width=85mm]{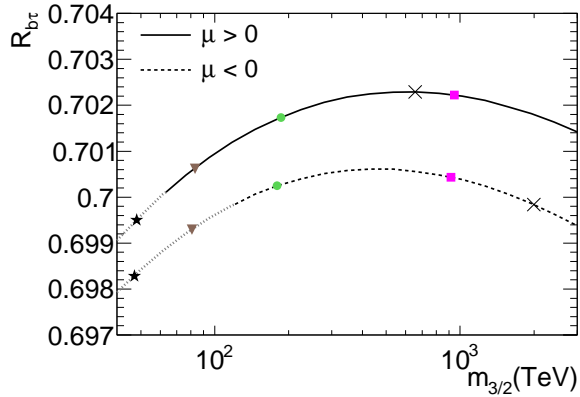}
  \end{center}
  \caption{$R_{b\tau}$ as a function of $m_{3/2}$, taking $m_{\bf 10}
    = m_{\bf \bar{5}} = m_{3/2}$, and $m_{H {\bf 5}} = m_{H {\bf
        \bar{5}}} = 0.8m_{3/2}$.  The upper and lower lines are for
    $\mu > 0$ and $\mu < 0$, respectively. The light gray (dotted)
    part of the lines correspond to the regions excluded by the Wino
    or gluino mass bounds.  The cross mark on each line shows the
    point with $m_{\tilde{W}}=3.0\ {\rm TeV}$.  The other marks on
    the lines show the points with $\tan\beta$ = 4.0 ($\bigstar$), 3.5
    ($\blacktriangledown$), 3.0 ({\Large $\bullet $}), and
    2.4($\blacksquare$).}
  \label{fig:universal}
\end{figure}

In order to see how $R_{b\tau}$ depends on the masses of SUSY
particles, we calculate $R_{b\tau}$ by taking $m_{\bf \bar{5}}=m_{\bf
  10}$, and $m_{H{\bf 5}}=m_{H{\bf \bar{5}}}=0.8m_{\bf 10}$.\footnote
{In order for the successful electroweak symmetry breaking with
  $m_{H_u}^2 (M_{\rm S})<0$, $m_{H{\bf 5}}^2$ is preferred to be
  smaller than sfermion masses in the present scenario.}
Then, with $\tan\beta$ being fixed, only one free parameter remains,
which is chosen to be $m_{3/2}$.  In Fig.\ \ref{fig:MgravVsR},
$R_{b\tau}$ is plotted as a function of $m_{3/2}$ for several values
of $\tan\beta$.  We also show the ratio of $m_{\bf 10}/m_{3/2}$ on
each line.  Some of the lines end at the middle of the figure.  This
is because, with $m_{\rm 10}^2$ being positive, the Higgs mass of
$\sim 125\ {\rm GeV}$ cannot be realized if the gravitino mass is too
large.  For $\mu>0$, $R_{b\tau}$ becomes enhanced with larger
$m_{3/2}$ or with larger $\tan\beta$; such a choice of parameters
makes $\Delta_b$ being negative and sizable, resulting in the
suppression of the bottom Yukawa coupling constant below $M_{\rm S}$.
For $\mu<0$, on the contrary, $R_{b\tau}$ becomes suppressed with
larger $m_{3/2}$ or larger $\tan\beta$ in the large $\tan\beta$ region
($\tan\beta\gtrsim 10$) since the sign of $\Delta_b$ is positive in
this case.  In the low $\tan\beta$ region ($\tan\beta\lesssim 10$)
with $\mu<0$, this is not the case since the heavy Higgs contributions
to $\Delta_b$, whose sign is uncorrelated to the sign of $\mu$, become
comparable to the sbottom-gluino and stop-chargino contributions.  We
note here that, in Fig.\ \ref{fig:MgravVsR}, we consider the the
gravitino mass up to a few PeV, with which the Wino mass becomes $\sim
3\ {\rm TeV}$.  In such a parameter region, the neutral Wino is the
LSP and hence the thermal relic density of the Wino becomes comparable
to the dark matter density \cite{Hisano:2006nn}.

We can see that $R_{b\tau}\sim 0.7$ for both $\mu>0$ and $\mu<0$ when
the scalar masses are of the same order of $m_{3/2}$.  This fact
indicates that, in the AMSB/PGM scenario, the threshold correction at
the GUT scale needs to be sizable for successful Yukawa
unification. To make this point clearer, we calculate $R_{b\tau}$ for
the case where all the scalar masses are of the same order of
$m_{3/2}$.  In Fig.\ \ref{fig:universal}, we plot $R_{b\tau}$ as a
function of $m_{3/2}$, taking $m_{\bf \bar{5}}=m_{\bf 10}=m_{3/2}$,
and $m_{H{\bf 5}}=m_{H{\bf \bar{5}}}=0.8m_{3/2}$.  In this case,
masses of all the sfermions, including stops, are required to be much
heavier than $\sim 10\ {\rm TeV}$, and hence relatively small
$\tan\beta$ is needed.  (See Fig.\ \ref{fig:scan}.)  On each line, we
show the value of $\tan\beta$.  We can see that $R_{b\tau}\sim 0.7$
with such a choice of parameters; this is because $\Delta_b$ is
suppressed due to the smallness of $\tan\beta$.

So far, we have seen that a significant hierarchy between the scalar
masses and the gravitino mass is needed to make $R_{b\tau}$ close to
$1$.  If we require $|R_{b\tau}-1|<0.1$, for example, $m_{\bf 10}^2$
and $m_{\bf \bar{5}}^2$ are required to be of $O(1)\ \%$ of
$m_{3/2}^2$.  Because the supergravity effects are expected to make
the SUSY breaking mass squared parameters to be of $O(m_{3/2}^2)$,
such a choice of $m_{\bf 10}^2$ and $m_{\bf \bar{5}}^2$ require the
tuning of the parameters in the K\"ahler potential at the level of
$O(1)\ \%$.

\begin{figure}[t]
  \begin{minipage}{0.48\hsize}
    \begin{center}
      \includegraphics[width=85mm]{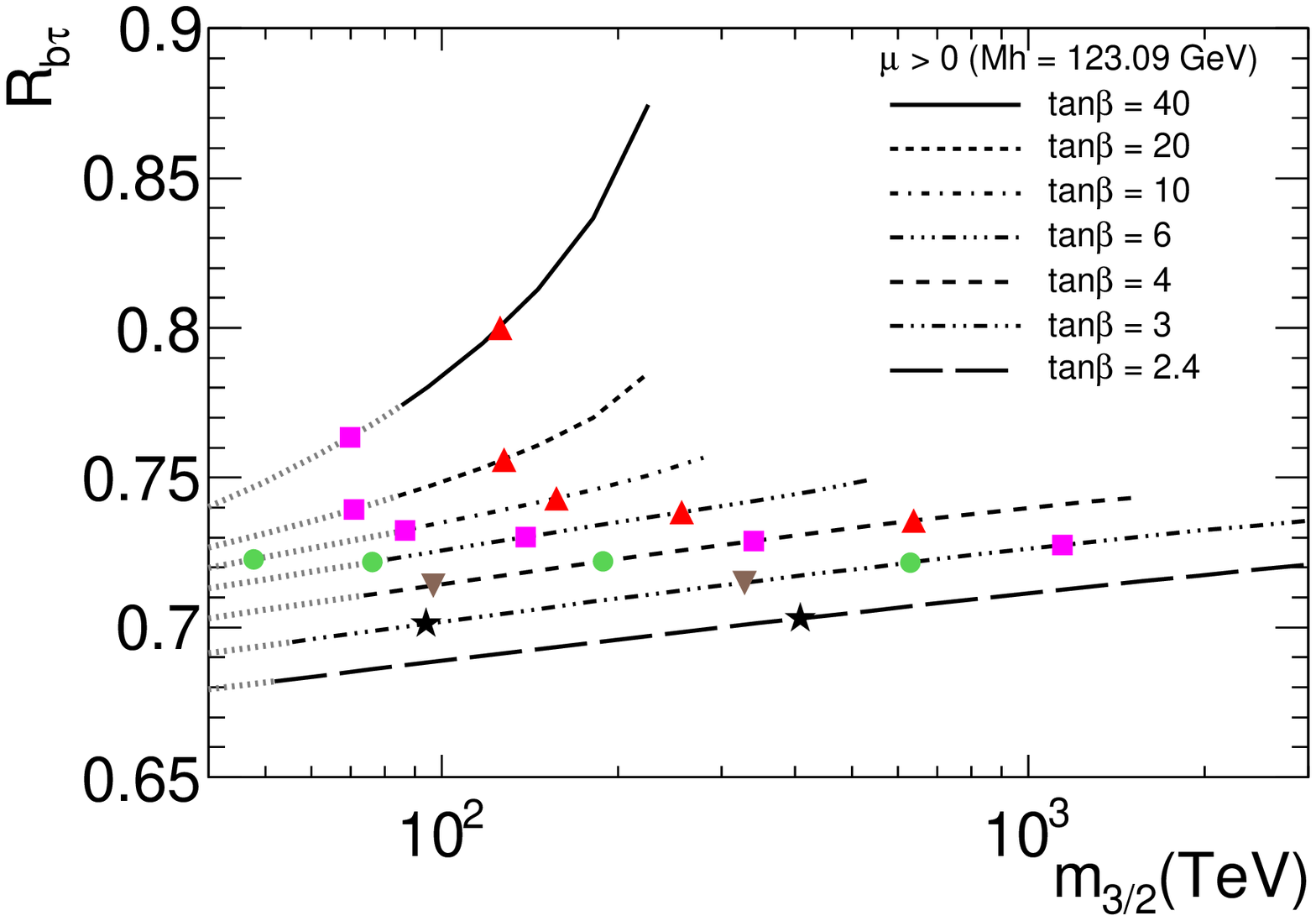}
    \end{center}
  \end{minipage}
  \begin{minipage}{0.48\hsize}
    \begin{center}
      \includegraphics[width=85mm]{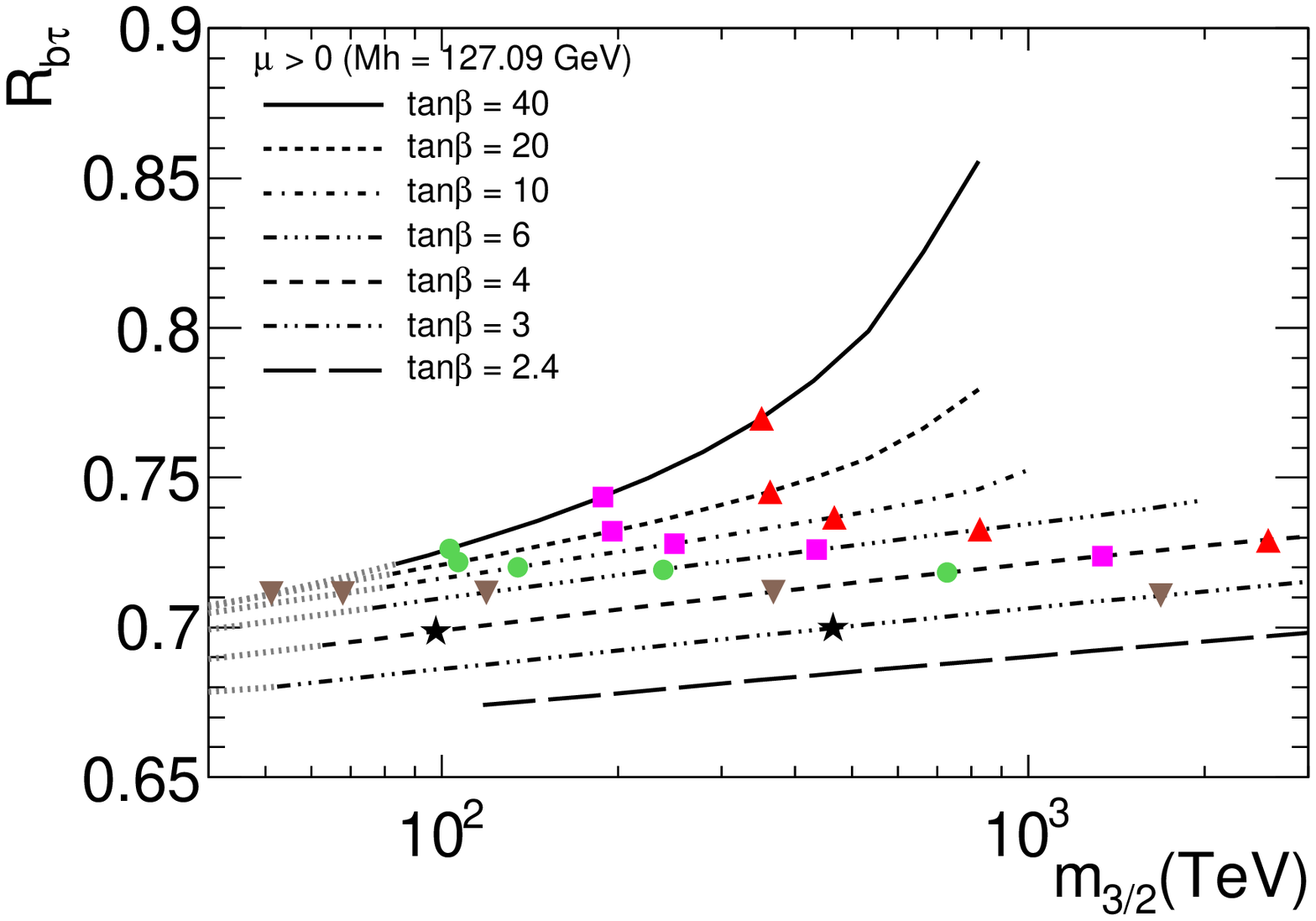}
    \end{center}
  \end{minipage}
  \caption{Same as Fig.\ \ref{fig:MgravVsR} for $\mu>0$, except for
    $m_h=123.09\ {\rm GeV}$ (left) and $127.09\ {\rm GeV}$ (right).  }
  \label{fig:MgravVsR_HiggsMassDep}
\end{figure}

We would also like to comment on the effects of the uncertainty in the
Higgs mass.  Although, experimentally, the Higgs mass is determined
with the accuracy of $0.21\ {\rm GeV}$, it is expected that the
theoretical calculation of the Higgs mass has larger uncertainty of a
few GeV.  In order to see how our results change with the variation of
the Higgs mass, we calculate $R_{b\tau}$ using $m_h=123.09\ {\rm GeV}$
and $127.09\ {\rm GeV}$, taking $m_{\bf \bar{5}} = m_{\bf 10}$, $m_{H
  {\bf 5}} = m_{H {\bf \bar{5}}} = 0.8m_{\bf 10}$, and $\mu >0$.  The
results are shown in Fig.\ \ref{fig:MgravVsR_HiggsMassDep}.  Comparing
with Fig.\ \ref{fig:MgravVsR}, we can see that, even if we vary the
threoretical prediction of the Higgs mass within the theoretical
uncertainty, $R_{b\tau}\sim 0.7$ when the gluino masses are much
smaller than the sfermion masses.  Thus, our main results are
unchanged.

\begin{figure}
  \begin{tabular}{cc}
    \begin{minipage}[t]{0.49\hsize}
      \centering
      \includegraphics[width=90mm]{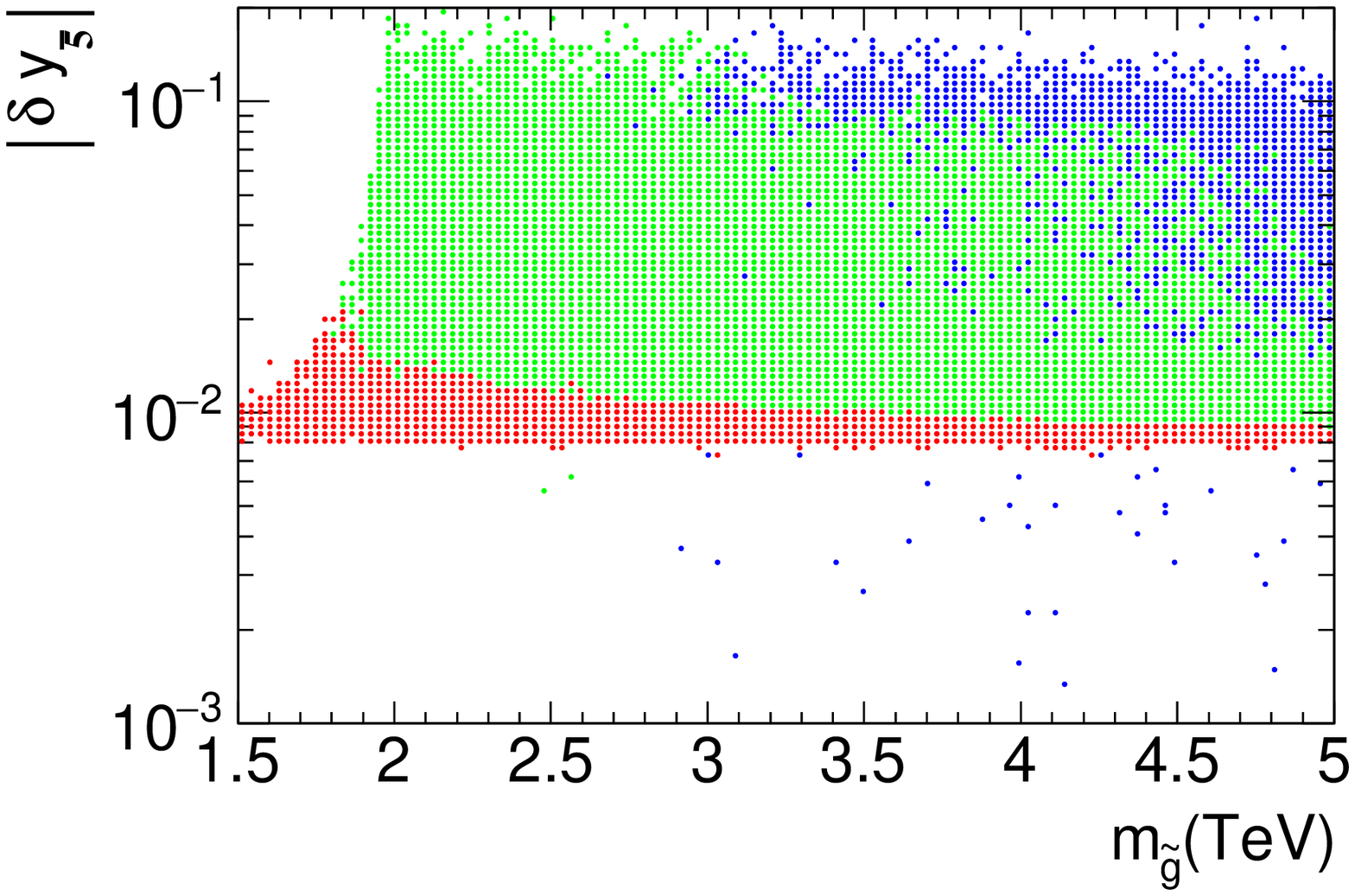}
    \end{minipage} &
    \begin{minipage}[t]{0.49\hsize}
      \centering
      \includegraphics[width=90mm]{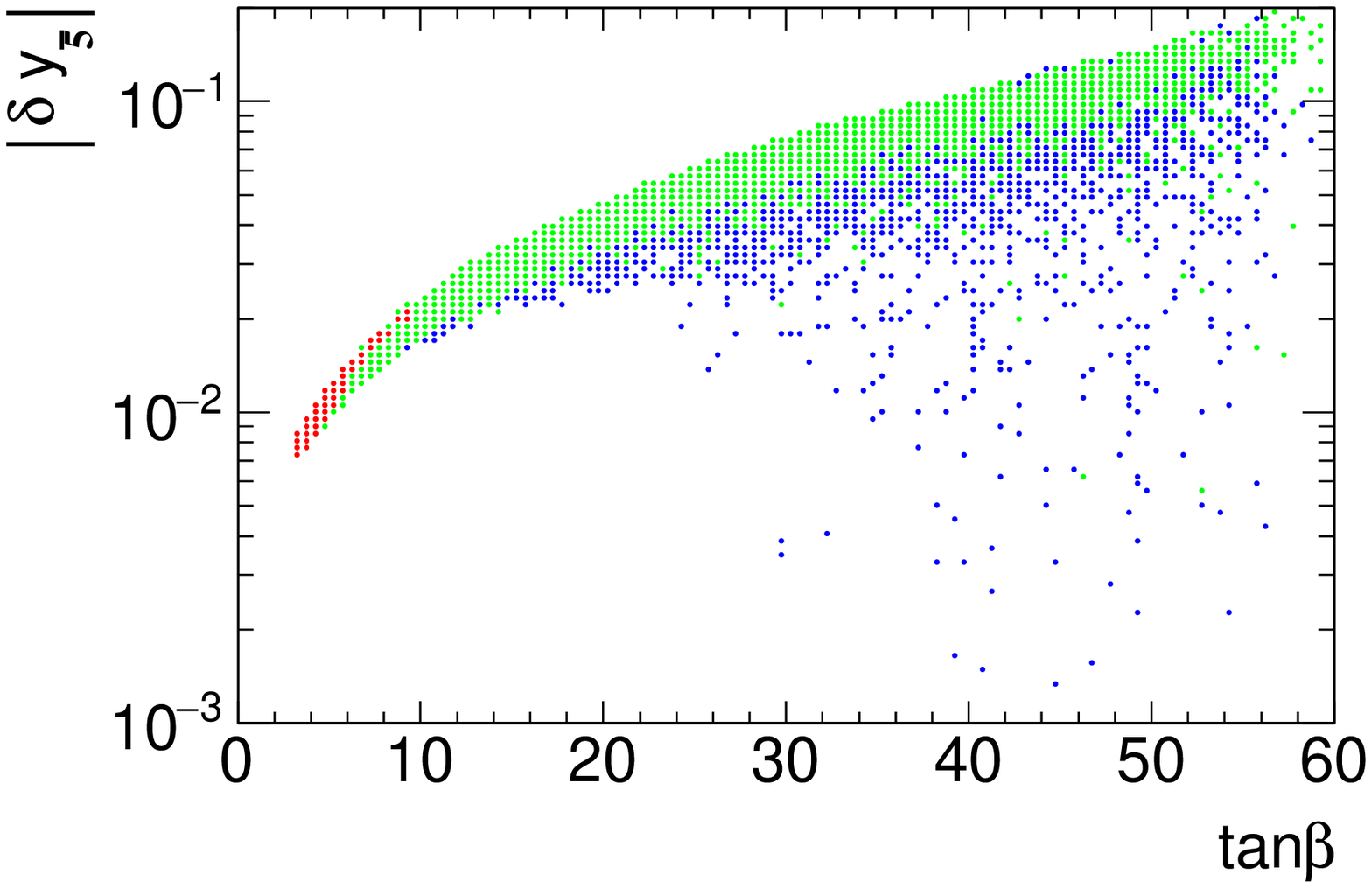}
    \end{minipage}\\
    \begin{minipage}[t]{0.49\hsize}
      \centering
      \includegraphics[width=90mm]{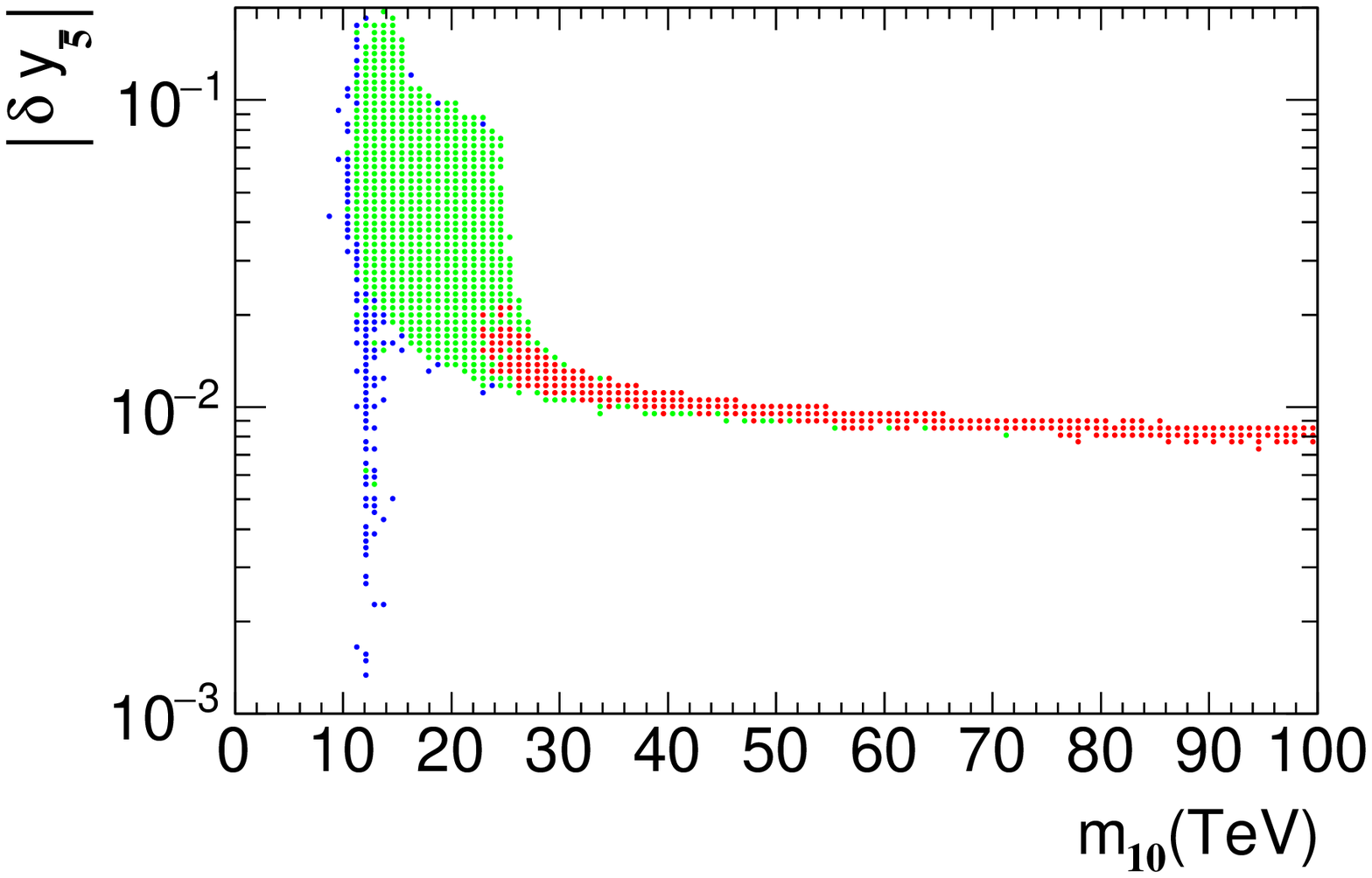}
    \end{minipage} &
    \begin{minipage}[t]{0.49\hsize}
      \centering
      \includegraphics[width=90mm]{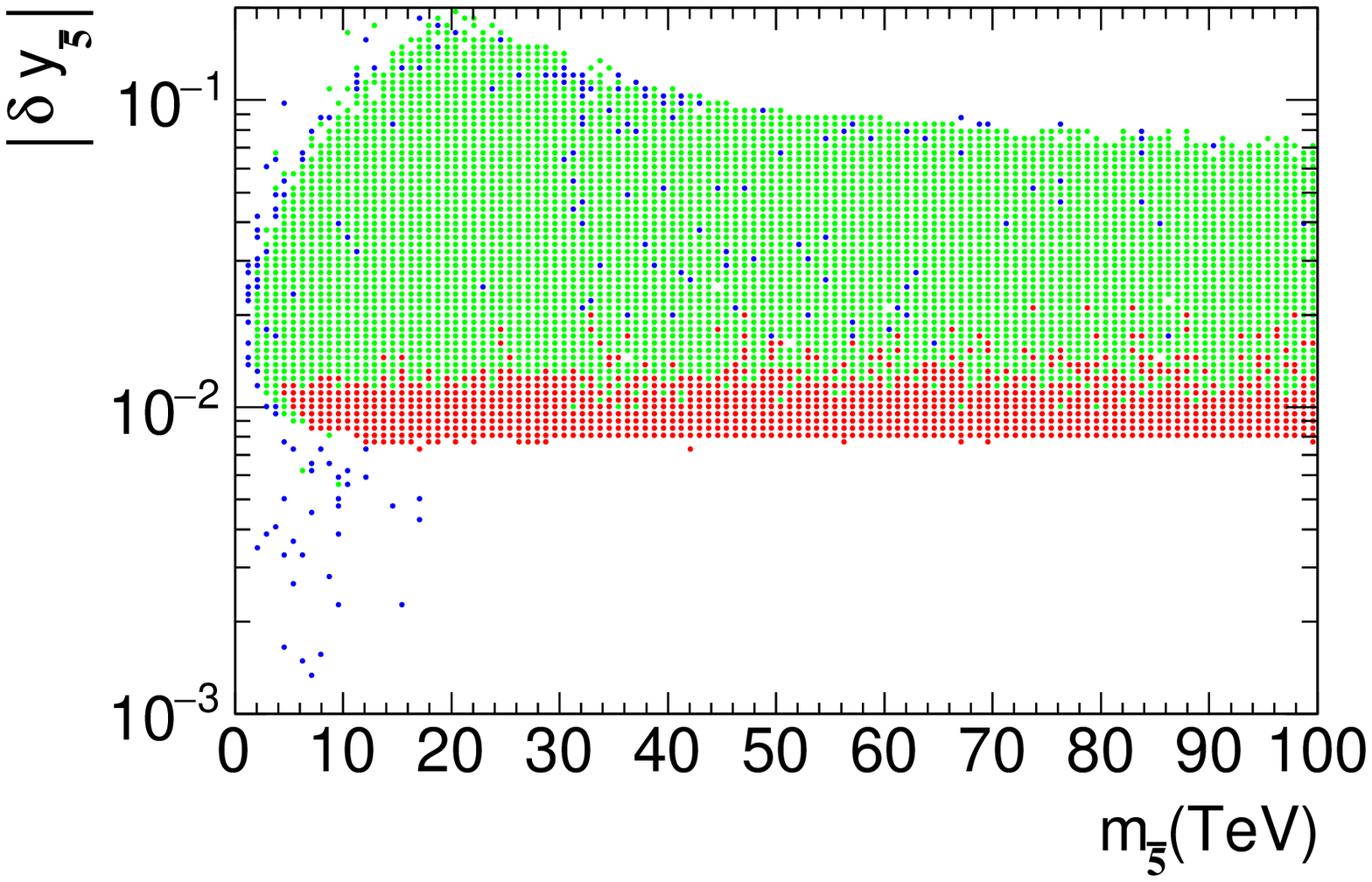}
    \end{minipage}\\
    \begin{minipage}[t]{0.49\hsize}
      \centering
      \includegraphics[width=90mm]{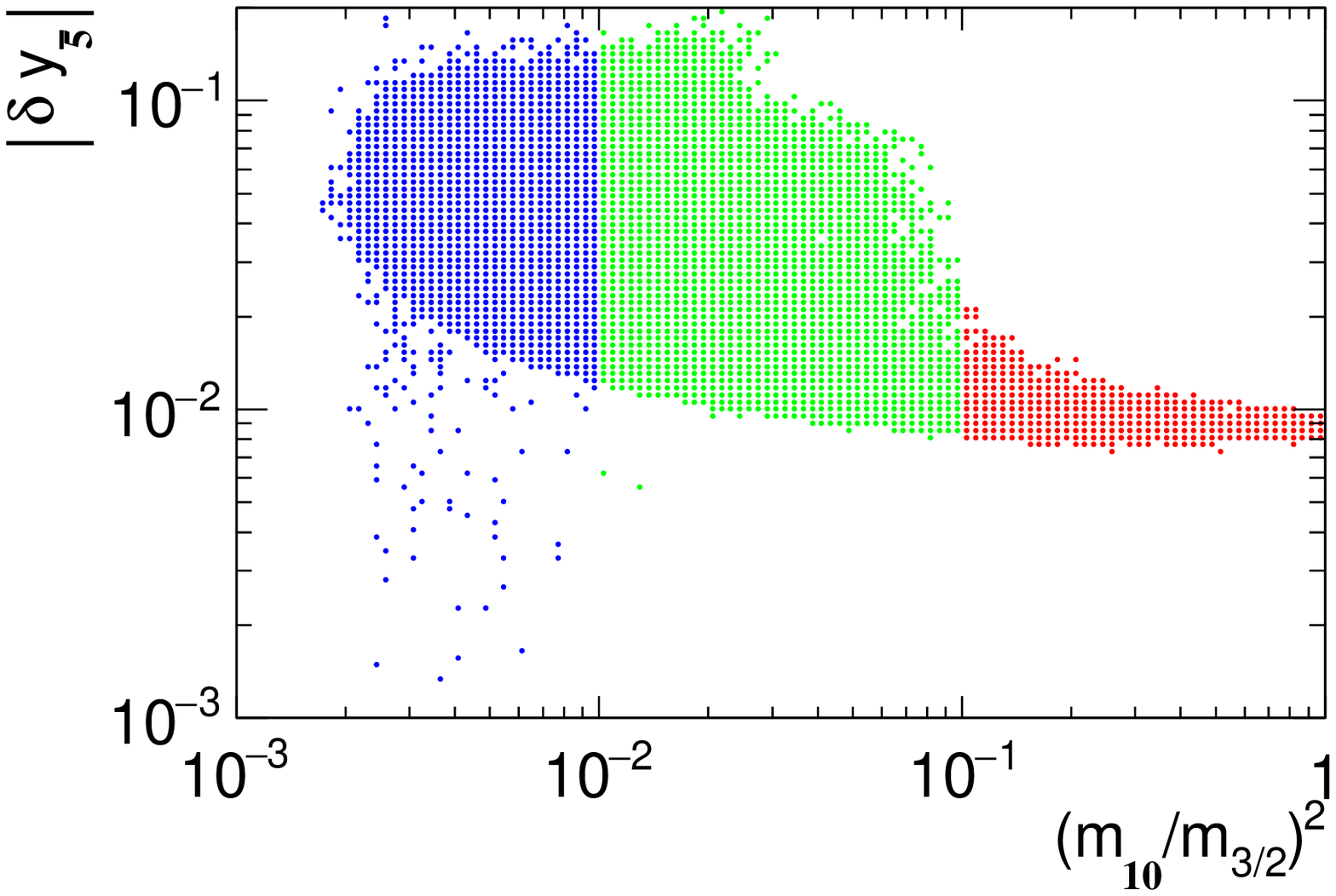}
    \end{minipage} &
    \begin{minipage}[t]{0.49\hsize}
      \centering
      \includegraphics[width=90mm]{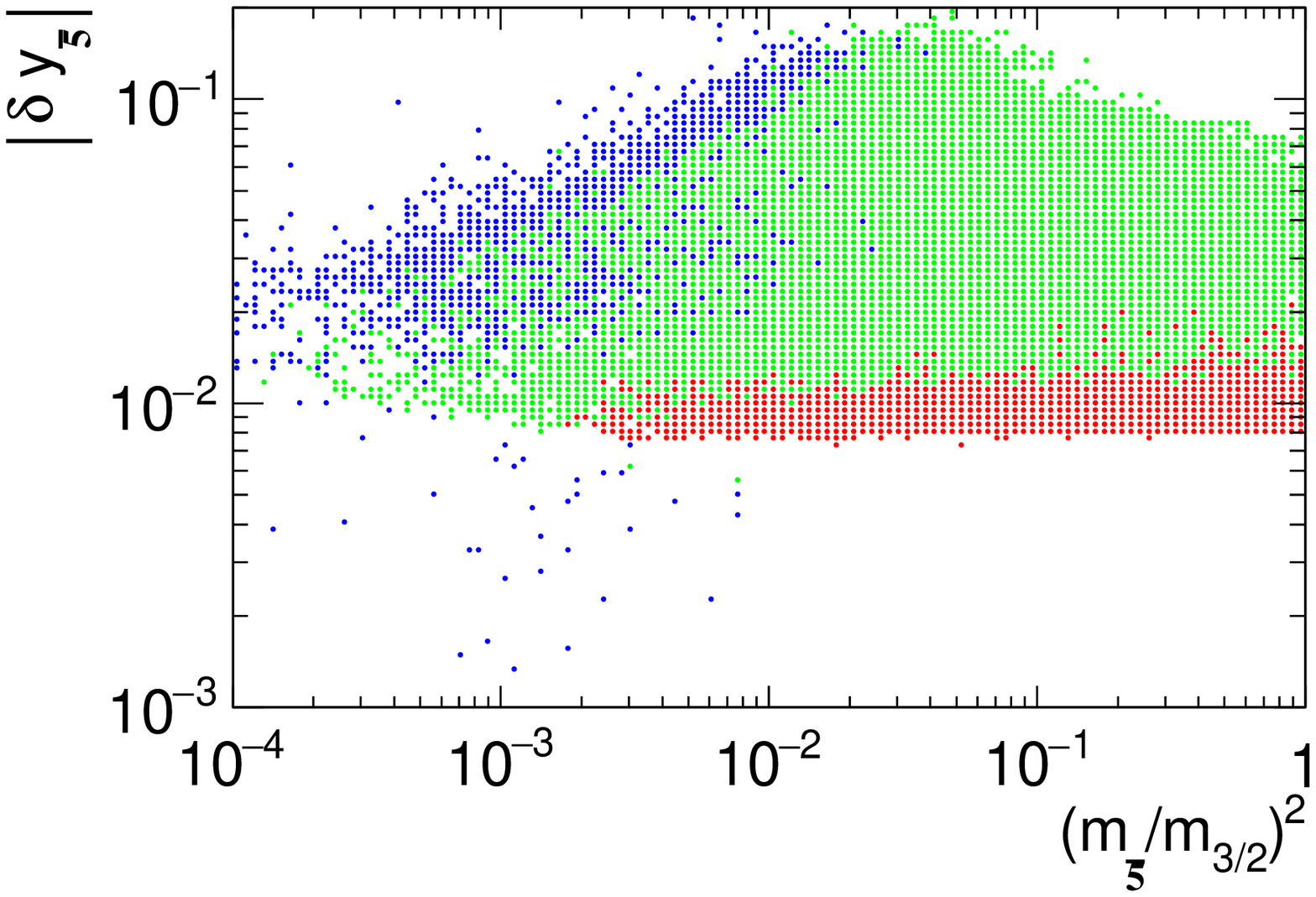}
    \end{minipage}\\
  \end{tabular}
  \caption{The distribution of the sample points on $m_{\tilde{g}}$ vs.\ 
    $|\delta y_{\bf \bar{5}}|$, $\tan\beta$ vs.\ $|\delta y_{\bf \bar{5}}|$,
    $m_{\bf 10}$ vs.\ $|\delta y_{\bf \bar{5}}|$, $m_{\bf \bar{5}}$ vs.\
    $|\delta y_{\bf \bar{5}}|$, $(m_{\bf 10} / m_{3/2})^2$ vs.\ $|\delta
    y_{\bf \bar{5}}|$, and $(m_{\bf \bar{5}} / m_{3/2})^2$ vs.\ $|\delta
    y_{\bf \bar{5}}|$ planes.  The meanings of the colors are the same
    as those in Fig.\ \ref{fig:Rbtau}.}
  \label{fig:dy}
\end{figure}

Before closing this section, we also study the difference between
$y_b(M_{\rm GUT})$ and $y_\tau (M_{\rm GUT})$:
\begin{align}
  \delta y_{\bf \bar{5}} \equiv
  y_b (M_{\rm GUT}) - y_\tau (M_{\rm GUT}).
\end{align}
In order to see how large $\delta y_{\bf \bar{5}}$ is in the AMSB/PGM
scenario, in Fig.\ \ref{fig:dy}, we show the distribution of $|\delta
y_{\bf \bar{5}}|$ as a result of our scan.  As one can see, $|\delta
y_{\bf \bar{5}}|$ becomes smaller as $\tan\beta$ decreases; this is
due to the smallness of $y_b$ and $y_\tau$ in the region with
relatively small $\tan\beta$.  Implication of this will be discussed
in the next section.

\section{Implications}
\label{sec:implications}
\setcounter{equation}{0}

Let us now discuss implications of our numerical results.  In
particular, we consider how small $|R_{b\tau}-1|$ should be in order
for the successful $b$-$\tau$ unification.  If $y_b(M_{\rm GUT})\neq
y_\tau (M_{\rm GUT})$, the difference is expected to be compensated by
corrections at the GUT scale.  The possible size of the corrections at
the GUT scale is strongly model-dependent.  

Due to the mass splitting of the particles at the GUT scale,
$R_{b\tau}$ may deviate from $1$.  We expect that the threshold
correction due to such a mass splitting is estimated as
\begin{align}
  \delta y_f \sim
  \beta_{y_f} \log \frac{M_{\rm GUT}+\delta M_{\rm GUT}}{M_{\rm GUT}},
\end{align}
with $f=b$ and $\tau$, where $\beta_{y_f}$ denotes the
$\beta$-function of $y_f$, and $\delta M_{\rm GUT}$ is the typical
size of the mass splitting of the GUT-scale particles.  As far as
$\delta M_{\rm GUT}\sim O(M_{\rm GUT})$, such an effect results in
$|R_{b\tau}-1|$ of $O(10^{-2})$, because $\beta_{y_f}$ is at most of
the order of $y_f/16\pi^2$, and hence $R_{b\tau}\sim 0.7$ is hardly
explained by this effect.

Another class of correction may come from the effective operators
containing the fields which is responsible for the breaking of the GUT
symmetry.  Schematically, the superpotential responsible for such a
correction, which is dimension-$5$ or higher, can be written as
\begin{align}
  W_{\rm Higher\,Dim.} =
  \frac{c}{M_*} \Sigma T \bar{F} \bar{H},
  \label{W(HigherDim)}
\end{align}
where $M_*$ is the mass scale of the mechanism generating $W_{\rm
  Higher\,Dim.}$, while $c$ is determined by the coupling constants in
the model.  Here, $\bar{F}$ and $T$ are superfield in ${\bf \bar{5}}$
and ${\bf 10}$ representations of $SU(5)$, which contain third
generation quarks and leptons, and $\bar{H}$ is the superfield in
${\bf \bar{5}}$ representation containing down-type Higgs.  In
addition, $\Sigma$ is the field responsible for the breaking of
$SU(5)\rightarrow SU(3)_C\times SU(2)_L\times U(1)_Y$; hereafter, to
make our points clearer, $\Sigma$ is assumed to be in the adjoint
representation of $SU(5)$.

The superpotential of the form of Eq.\ \eqref{W(HigherDim)} may arise
from an unknown non-perturbative dynamics at the cut-off scale (like
the Planck or string scale, identifying $M_*$ as a cut-off scale), or
by integrating out particles whose masses are above the GUT scale.
Here, let consider a simple example of the latter.  We introduce the
following superpotential:
\begin{align}
  W' = M_* F' \bar{F}'
  + \kappa \Sigma F' \bar{F}
  + y_{\bf \bar{5}}' T \bar{F}' \bar{H}
  + y_{\bf \bar{5}} T \bar{F} \bar{H},
  \label{W0}
\end{align}
where $F'$ and $\bar{F}'$ are new superfields in ${\bf 5}$ and ${\bf
  \bar{5}}$ representations, respectively; with $M_*\gtrsim M_{\rm
  GUT}$, the superpotential of the form of Eq.\ \eqref{W(HigherDim)}
is obtained after integrating out $F'$ and $\bar{F}'$.  We denote the
vacuum expectation value of $\Sigma$ as $\langle\Sigma\rangle={\rm
  diag}(2\sigma,2\sigma,2\sigma,-3\sigma,-3\sigma)$, assuming that
$\sigma\sim O(M_{\rm GUT})$.  Then, with the superpotential given in
Eq.\ \eqref{W0}, $b_R^{c}$ and $l_L$ are given by
\begin{align}
  b_R^{c} & = \bar{F}_3 \cos\theta_{b_R^{c}}
  + \bar{F}'_3 \sin\theta_{b_R^{c}},
  \\
  l_L &= \bar{F}_2 \cos\theta_{l_L}
  + \bar{F}'_2 \sin\theta_{l_L},
\end{align}
where $\bar{F}_3^{(\prime)}$ and $\bar{F}_2^{(\prime)}$ are upper
three and lower two components of $\bar{F}^{(\prime)}$, respectively,
and
\begin{align}
  \tan\theta_{b_R^{c}} = -\frac{2\kappa\sigma}{M_*},
  ~~~
  \tan\theta_{l_L} = \frac{3\kappa\sigma}{M_*}.
\end{align}
Then, the Yukawa coupling constants of $b$ and $\tau$ at the GUT scale
are estimated as
\begin{align}
  y_b (M_{\rm GUT}) &= y_{\bf \bar{5}} -2 \epsilon y_{\bf \bar{5}}'
  + O(\epsilon^2),
  \\
  y_\tau (M_{\rm GUT}) &= y_{\bf \bar{5}} + 3 \epsilon y_{\bf \bar{5}}'
  + O(\epsilon^2),
\end{align}
with
\begin{align}
  \epsilon \equiv \frac{\kappa \sigma}{M_*},
\end{align}
and hence
\begin{align}
  R_{b\tau} = 1 - 5 \epsilon \frac{y_{\bf \bar{5}}'}{y_{\bf \bar{5}}}
  + O(\epsilon^2).
\end{align}
In addition, 
\begin{align}
  \delta y_{\bf \bar{5}} \simeq 
  - 5 \epsilon y_{\bf \bar{5}}' 
  + O(\epsilon^2).
\end{align}

$R_{b\tau}$ may significantly deviate from $1$ in this set up.  If
$y_{\bf \bar{5}}'\sim y_{\bf \bar{5}}$, $|R_{b\tau}-1|\sim O(0.1)$
requires $\epsilon\sim O(0.1)$.  In such a case, the quarks and
leptons, which are embedded into the same $SU(5)$ multiplet in the
simplest scenario, are given by different admixture of the fields at
the GUT scale or above.  On the contrary, for $y_{\bf \bar{5}}'\gg
y_{\bf \bar{5}}$, $|R_{b\tau}-1|\sim O(0.1)$ is possible even with
$\epsilon\ll O(0.1)$.  In particular, when $\tan\beta$ is not so
large, $y_b$ and $y_\tau$ are much smaller than $1$ and hence $\delta
y_{\bf \bar{5}}\ll 1$ (see Fig.\ \ref{fig:dy}).  Then, the $b$-$\tau$
unification may be realized with $M_*$ much larger than the GUT scale
(like $M_*$ as large as the Planck scale).  In such a case, however,
the quarks and leptons have new Yukawa interactions much stronger than
those in the MSSM, which may introduce new flavor and CP problems in
SUSY model.  In particular, it is unclear if the new field $F'$
selectively couples to the third generation quarks and leptons.  If
the coupling between $F'$ and first or second generation quarks and
leptons is as strong as that to third generation ones, the hierarchy
of the SM Yukawa coupling constants are easily spoiled.  We also note
here that, if non-trivial flavor mixings or CP violations exist in
such new couplings, they may affect the SUSY breaking scalar mass
squared parameters via the renormalization group runnings
\cite{Hall:1985dx}.  Such an effect may give sizable contributions to
low energy flavor and CP violating observables.

\section{Summary}
\label{sec:summary}
\setcounter{equation}{0}

We have studied the $b$-$\tau$ unification in SUSY model with the
AMSB/PGM mass spectrum.  In the model of our interest, sfermions as
well as Higgsinos acquire masses from direct interactions with SUSY
breaking fields while gaugino masses are from AMSB effect.
Consequently, the gaugino masses (as well as the SUSY breaking
tri-linear scalar coupling constants) are one-loop suppressed compared
to the sfermion masses.  In order for the accurate study of the
renormalization group effects on coupling constants, we have
considered three different effective theories, i.e., SM,
$\tilde{G}$SM, and MSSM.  We have used a numerical program in which
the two-loop RGEs in these effective theories, as well as threshold
corrections at the matching scales, are implemented, and calculated
the Yukawa coupling constants at the GUT scale.  In order to
understand the viability of the Yukawa unification in the AMSB/PGM
scenario, we have performed the parameter scan and calculated
$y_b(M_{\rm GUT})$ and $y_\tau (M_{\rm GUT})$ for about $5\times
10^{4}$ sample points.

We have found that the naive mass spectrum of the AMSB/PGM scenario,
in which the sfermion masses are of the order of the gravitino mass,
predicts $y_b(M_{\rm GUT})\sim 0.7 y_\tau(M_{\rm GUT})$, which
conflicts with the $b$-$\tau$ Yukawa unification in the simple set up.
In order to solve this discrepancy, one may consider sizable
corrections at the GUT scale.  In such a case, a non-trivial flavor
structure is suggested at the GUT scale, which may affect low-energy
flavor and CP violating observables.  Another resolution may be to
adopt suppressed sfermion masses compared to the gravitino mass.  As a
result of our parameter scan, we found sample points with
$|R_{b\tau}-1|<0.1$, for example, when the sfermion mass squared
parameters, $m_{\bf 10}^2$ and $m_{\bf \bar{5}}^2$, are of $O(1)\ \%$
of $m_{3/2}^2$.  Because the expectation is that $m_{\bf 10}^2$ and
$m_{\bf \bar{5}}^2$ are of $O(m_{3/2}^2)$, this may suggest the $O(1)\
\%$ level tuning of the parameters in the K\"ahler potential to
suppress the scalar masses.

\vspace{5mm}
\noindent {\it Acknowledgements}: The work of T.M. is supported by
JSPS KAKENHI No.\ 26400239.

\end{document}